\documentclass[9pt,journal]{IEEEtran}
\usepackage{float}
\usepackage{comment}
\usepackage[table]{xcolor}
\usepackage{cite}      
\usepackage{graphicx}  
\usepackage{psfrag}    
\usepackage{placeins}
\usepackage{url}
\usepackage{stfloats}
\usepackage{times}
\usepackage{textcomp}
\usepackage{amsmath}   
\usepackage{amsfonts,amsthm,amssymb}
\usepackage{mathtools}
\usepackage{nicefrac}
\usepackage[none]{hyphenat}
\usepackage{balance}
\usepackage{widetext}
\usepackage{algpseudocode}
\usepackage{algorithm}
\usepackage[caption=false,font=footnotesize]{subfig}

\usepackage{cite}

\usepackage{array}
\usepackage{fontenc}
\usepackage{color}
\usepackage{etoolbox}
\usepackage{bm}

\begin{document}
\title{Polarimetric Radar Cross-Sections of Pedestrians at Automotive Radar Frequencies}
\author{Yoshana~Deep$^{\dagger}$, Patrick~Held$^{\ddagger}$, Shobha~Sundar~Ram$^{\dagger}$, Dagmar Steinhauser$^{\ddagger}$, Anshu Gupta$^{\ast}$, Frank Gruson$^{\ast}$, Andreas Koch$^{\ast}$, Anirban Roy$^{\ast}$
\thanks{Y. D.$^{\dagger}$ and S.S.R.$^{\dagger}$ are with the Indraprastha Institute of Information Technology Delhi, New Delhi 110020 India. E-mail: \{yoshana17128,shobha\}@iiitd.ac.in.}%
\thanks{P. H.$^{\ddagger}$ and D. S.$^{\ddagger}$ are with CARISSMA, Technische Hochschule Ingolstadt. E-mail:\{Patrick.Held,Dagmar.Steinhauser\}@carissma.eu .}
\thanks{A.G.$^{\ast}$, F.G.$^{\ast}$, A.K.$^{\ast}$ and A. R.$^{\ast}$ are with Continental, Business Unit ADAS, Peter-Dornier-Straße 10, Lindau, Germany. E-mail:\{anshu.gupta,frank.gruson,andreas.6.koch,anirban.roy\} @continental-corporation.com}
}
\maketitle

\begin{abstract}
Simulation of radar cross-sections (RCS) of pedestrians at automotive radar frequencies forms a key tool for software verification test beds for advanced driver assistance systems. Two commonly used simulation methods are: the computationally simple scattering center model of dynamic humans; and the shooting and bouncing ray technique based on geometric optics. The latter technique is more accurate but due to its computational complexity, it is usually used only for modeling scattered returns of still human poses. In this work, we combine the two methods in a linear regression framework to accurately estimate the scattering coefficients or reflectivies of point scatterers in a realistic automotive radar signal model which we subsequently use to simulate range-time, Doppler-time and range-Doppler radar signatures. The simulated signatures show a normalized mean square error below $10\%$ and a structural similarity above $81\%$ with respect to measurement results generated with an automotive radar at 77 GHz.
\end{abstract}
\maketitle
\section{Introduction}
\label{Sec:Intro}
Pedestrians - especially children, senior citizens and those with disabilities - are among the most vulnerable road users. Anywhere from 12\% to 38\% of the road fatalities occur to pedestrians \cite{world2017more}. Recently, there has been significant research focus on developing advanced driver assistance systems (ADAS) for improving driving conditions and reducing road fatalities. Pedestrian detection, one of the key objectives of ADAS, has been researched with both automotive cameras \cite{andrade2018novel,silberstein2014vision} and radars \cite{belyaev2013use,engels2017advances}. Camera images offer key features - in the form of shapes, sizes and texture cues - for enabling automatic detection and recognition. However, the performance of the camera is affected by light and visibility conditions. Automotive radars, unlike cameras, can operate continuously, under low visibility conditions and, in some cases, in non-line-of-sight conditions as well. Most importantly, the swinging motions of a pedestrian's arms and legs, while walking, give rise to distinctive Doppler radar signatures \cite{ghaleb2008micro,kim2009human,kim2015human,chen2002time,geisheimer2002high,van2010human,ram2007analysis}. These micro-Doppler signals are different from those generated by other dynamic bodies on the road such as bicycles and cars and hence can be used for automatic target recognition \cite{ghaleb2008micro,rohling2010pedestrian,stolz2017multi,angelov2018practical,lee2017human,prophet2018pedestrian,mishra2019doppler}. 

The performances of these algorithms rely on the availability of large training databases gathered in a variety of scenarios. They must comprise of data from pedestrians of different ages, heights and girth; performing different activities and moving at different orientations with respect to the radar. There are two methods of generating the training data. One method is to collect the data from real pedestrians using actual automotive radar sensors. The advantage is that the training data is real and can be gathered both in laboratory conditions and during test drives. However, the disadvantage is that the database must be updated based on hardware modifications to the sensor or due to software changes in the signal processing. Second, the data may be corrupted by the presence of clutter from the local environment (both static and dynamic) and limitations of the sensor. Labelling of radar measurement data gathered over long test drives also requires painstaking efforts. Finally, pedestrians are dielectric bodies of much smaller RCS than other road targets and are unpredictable in terms of motion and posture. The alternative is to simulate the radar signatures \cite{ram2010simulation}. The advantage is that the simulated radar signatures can be rapidly generated for a variety of sensor parameters and target scenarios. Also, the simulations can be easily integrated with the radar test bed and signal processing platforms for rapid prototyping and validations. Finally, since the simulated data may be made free of channel artifacts such as clutter, the simulation results may facilitate identifying cause and effect of the underlying radar phenomenology. 

Simulations of radar micro-Doppler signatures have been extensively researched over the last decade \cite{ram2010simulation,fairchild2014classification,ren2018short,park2014simulation}. The methods have included simple pendulum models of the human motions \cite{smith2008multistatic,tsai2010real}; analytical models of walking motion derived from bio-mechanical experiments \cite{boulic1990global}; and computer animation models for describing more complex human motions \cite{ram2008microdoppler, erol2015kinect,poser,chattopadhyay2007human}. The motion models are subsequently combined with electromagnetic models of radar scattering off humans. \emph{Full wave electromagnetic solvers} yield very accurate predictions of radar cross-sections (RCS). However, they are not used for modeling humans due to the considerable computational complexity (in terms of time and memory) in modeling three-dimensional spatially large dielectric bodies at automotive radar frequencies (24 GHz and 77 GHz). Further, humans are dynamic and have a distinct pose and posture during each instant of any motion such as walking. A slightly less computationally expensive alternative is based on \emph{shooting and bouncing rays and geometric optics} and has been used for predicting the RCS of still humans at X-band and Ku-band frequencies \cite{dogaru2007computer, dogaru2008validation,le2007numerical}. However, the technique still remains computationally expensive and cannot be used to generate radar data at the pulse repetition frequencies typically used in automotive radars. Hence, ray tracing results cannot be directly used for generating radar signatures - such as high range-resolution profiles or Doppler-time spectrograms of humans - which provide key information for automatic target recognition.  
A third technique based on the \emph{point scattering center model} has been widely adopted for obtaining radar signatures of humans due to its low computational complexity \cite{van2003human,ram2010simulation}. Here the human is modelled as an extended target with multiple point scatterers. The scattering coefficient of each of these point scatterers has traditionally been determined from an approximated analytical expression for RCS of a primitive shape resembling the human body part corresponding to the point scatterer. The time-varying positions of the point scatterers are obtained from computer animation data. The resulting radar signatures have shown excellent correlation in terms of their micro-Doppler and micro-range features to the signatures derived from real measurement data. However, the method is very inaccurate in estimating the RCS magnitude due to the approximate nature of the primitive based model and because the model does not include the effects of shadowing and multipath interactions between the different body parts. The accurate estimation of RCS is, however, important for the implementation of radar detectors for generating the receiver operating curves. 

In this work, we propose a method for accurately predicting the RCS of pedestrians by combining electromagnetic ray tracing with the point scatterer model. Highly accurate estimates of RCS of the human are generated at video frame rate using the ray tracing technique. The reflectivities of the point scatterer human model are then estimated from the ray tracing RCS values using linear regression. These reflectivities are subsequently integrated with the scattering center model to generate the RCS at high radar sampling frequencies. Our method is founded on the assumption that since humans are slow moving targets, their scattering coefficients fluctuate slowly across multiple radar coherent processing intervals while the positions of the point scatterers change rapidly across multiple pulse repetition intervals. The proposed method, thereby, combines the advantages of high accuracy of ray tracing with the computational performance of scattering center modeling. We derive three types of radar signatures - high range-resolution profile, Doppler-time spectrogram and range-Doppler ambiguity diagram from the simulated data. We compare the signatures with similar signatures derived from measurement radar data at 77 GHz. Our results show a low normalized mean square error (below $10\%$) and high structural similarity (above $81\%$) between the measured and simulated radar signatures. 
We also present calibrated monostatic and bistatic RCS of humans at multiple aspect angles and polarizations for both the automotive radar frequencies (24 GHz and 77 GHz). 

The paper is organized as follows. In the following section, we present the simulation methodology for hybridizing the swift point scatterer modeling technique with the accurate but computationally heavy ray tracing method. In section \ref{Sec:ExpSetUp} we describe the experimental set up for jointly collecting radar measurement data and motion capture (MoCap) data. Finally, we present the simulated radar signatures of the pedestrian and provide the qualitative (\ref{Sec:RayTracingResults}) and quantitative (\ref{Sec:ResultsSig}) comparison with measurement results.

\emph{Notation}: We use the following conventions in our notations. Scalar variables are written with small letters; vectors are denoted with overhead arrows; and matrices are written with bold face capital letters. 
\section{Simulation Methodology}
\label{sec:SimMethod}
\begin{figure*}[!ht]
    \includegraphics[width=0.9\textwidth]{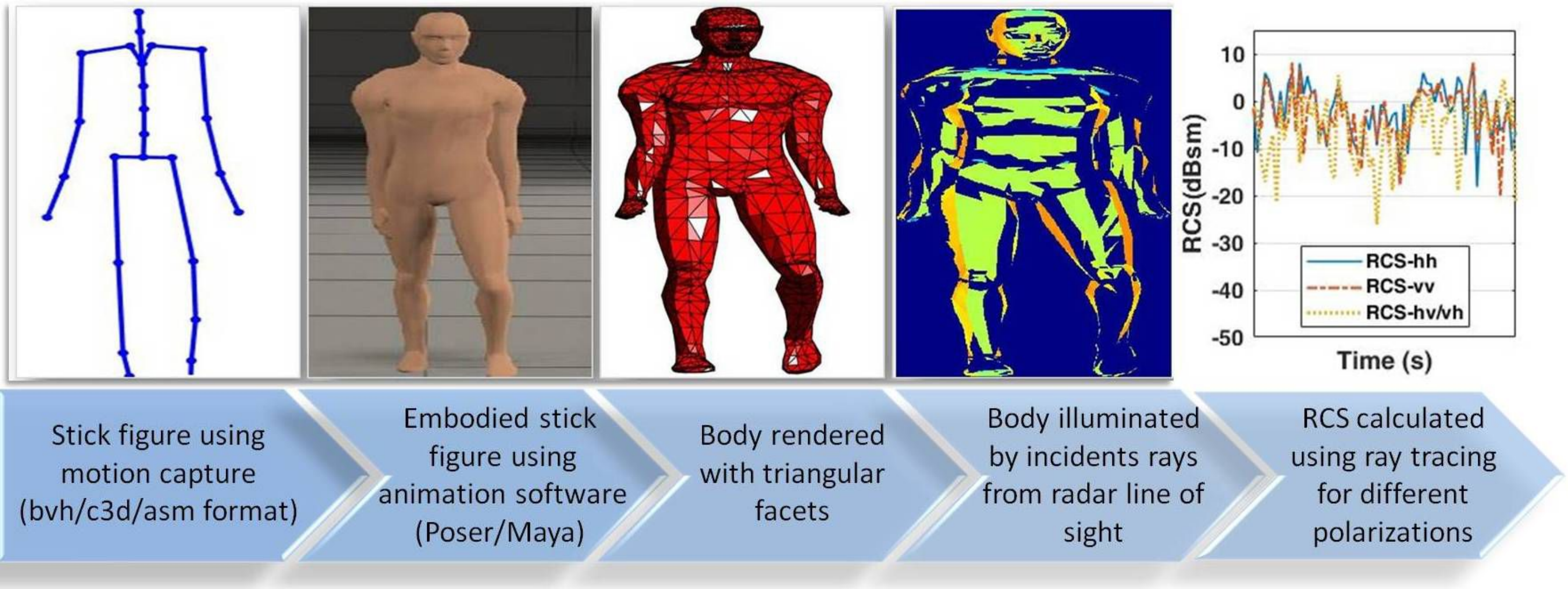}
    \caption{Motion capture (MoCap) data in stick figure format is embodied using animation software like Poser/Maya. Then the body is rendered with triangular facets. Radar cross-section is calculated using electromagnetic ray tracing for different polarizations.}
    \label{fig:Flowchart}
\end{figure*}

The objective of the proposed work is to simulate the radar scattered signal of dynamic humans in order to generate radar signatures such as range-time, Doppler-time and range-Doppler ambiguity plots. 
For our method, we rely on the availability of Motion Capture (MoCap) data of a dynamic human motion at video frame rate. 
We begin with a stick figure animated model of human motion obtained from MoCap technology as shown in Fig.~\ref{fig:Flowchart}. 
Each frame of MoCap data is exported to an animation software, such as Poser Pro from Smith Micro Software \cite{poser}, where the stick figure is embodied using one of the in-built libraries of an anatomically accurate human body. The human body is then rendered into a three-dimensional poly-mesh figure composed of triangular facets of suitable resolution \cite{garcia2010facet}. We consider the human body standing on an $x-y$ ground plane with the height along the positive $z$ axis.
We adapted the fairly standard shooting and bouncing ray technique proposed in \cite{LingHaoChouRiCheeandLee1989} on the poly-mesh human data for RCS estimation for different polarizations and aspect angles \cite{dogaru2007computer, dogaru2008validation}. In this technique, we consider a set of parallel, closely spaced illumination rays emanating from an incident plane at an incident angle $\phi^i$. Based on the interaction between the incident rays and the mesh triangles on the human body, we compute the scattered electric field along the direction of $\phi^s$. While there are several commercially available electromagnetic solvers that carry out ray tracing, we developed our solver in-house in order to customize it for speed and efficiency. The human body is a complex dielectric medium of skin, tissues and bone. However, at high frequencies (24 GHz and 77 GHz), there is little penetration through the skin and therefore, we model the human body as a single layer dielectric with relative permeability $\epsilon_r'(f_c) = \epsilon_r(f_c) + \frac{\sigma_{c}(f_c)}{j2 \pi f_c \epsilon_0}$. The dielectric constant and conductivity are $\epsilon_r = 6.63$ and $\sigma_{c} = 38.1$ S/m at 77 GHz  \cite{chen2014artificial} and $\epsilon_r = 50$ and $\sigma_{c} = 1$ S/m at 24 GHz \cite{dogaru2007computer,le2007numerical}. The scattered signal strength from the human is determined by the reflection coefficient of the dielectric surface.  
We estimate four types of RCS. They are the co-polarized horizontal ($\sigma^{\text{hh}}$) and vertical RCS ($\sigma^{\text{vv}}$) as well as the cross-polarized RCS ($\sigma^{\text{hv}}$ and $\sigma^{\text{vh}}$) at the video frame rate ($1/T_f$). Based on the incident and scattered angles, $\phi^{i}$ and $\phi^s$, we compute the bistatic RCS for different types of polarizations. When $\phi^s = \phi^{i}$, the RCS corresponds to the monostatic case.

In the following subsection (\ref{Sec:SimRadarModel}), we present our proposed method to use the ray tracing results to estimate scattering center coefficients which are subsequently used in a point scatterer model to generate scattered signals at suitable radar sampling frequencies.
\subsection{Proposed method: Estimation of scattering coefficients of point scatterer model using RCS from ray tracing}
\label{Sec:SimRadarModel}
The ray tracing method provides accurate estimates of the RCS of the whole human body based on the posture described by the MoCap data at video frame rate (30/48/60 Hz). However, the technique still remains computationally expensive and cannot be used to generate radar data at high radar sampling frequencies (of the order of GHz). Hence, ray tracing results cannot be directly used for generating radar signatures - such as range-time, Doppler-time and range-Doppler plots - of extended targets such as humans. In this section, we propose a method to obtain the radar signatures by hybridizing the ray tracing results and the scattering center model using the MoCap data.

\begin{figure}[!ht]
    \centering
    \includegraphics[width = 0.45\textwidth]{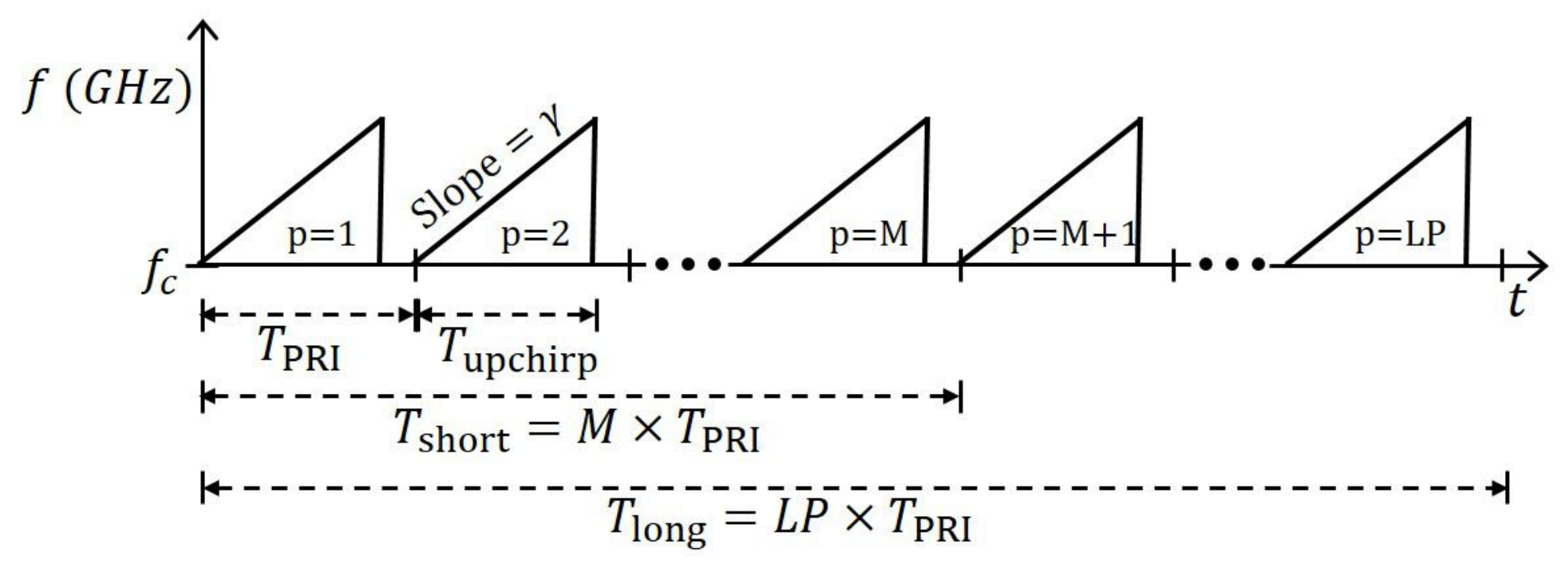}
    \caption{Radar signal model of linear frequency modulated continuous waveform of $T_{\text{upchirp}}$ duration with chirp rate $\gamma$ and $T_{\text{PRI}}$ pulse repetition interval (PRI). Each $T_{\text{upchirp}}$ consists of $N$ samples of $f_s = \frac{1}{T_s}$ sampling frequency. $T_{\text{long}}$ is the duration of $L$ coherent processing intervals (CPI) each consisting of $P$ PRIs. $T_{\text{short}}$ is the time interval between $M$ PRIs within $T_{\text{long}}$.}
    \label{fig:FMCW_waveform}
\end{figure}

We begin by assuming that a monostatic radar is located at the origin. We model the radar transmit waveform as a frequency modulated continuous waveform (FMCW) of center carrier frequency $f_c$, radar bandwidth ($BW$) and chirp factor ($\gamma =BW/T_{\text{upchirp}}$) as shown in Fig.~\ref{fig:FMCW_waveform}. The transmit signal, $x_p(\tau)$, over a single $p^{\text{th}}$ pulse repetition interval (PRI), $T_{\text{PRI}}$, is given by
\begin{gather}
\label{eq:TxWaveform}
    x_p(\tau) = \text{rect} \left(\frac{\tau}{T_{\text{PRI}}}\right) e^{j(2\pi f_c \tau + \pi\gamma \tau^2)},
\end{gather}
where
\begin{gather}
    \text{rect} \left(\frac{\tau}{T_{\text{PRI}}} \right) =
\begin{cases}
1,\,\, 0 \leq \tau \leq T_{\text{upchirp}}\\
0,\,\, T_{\text{upchirp}} < \tau < T_{\text{PRI}}.
\end{cases}
\end{gather}
The interval between the up chirp duration $T_{\text{upchirp}}$ and $T_{\text{PRI}}$ may be regarded as dead time. 
We assume that the transmitted waveform spans $T_{\text{long}}$ duration consisting of $L$ coherent processing intervals (CPI) each of $P$ PRIs. The radar signal falls upon a dynamic target of $B$ point scatterers with scattering coefficients or reflectivities, $\{a_b,\,\, b=1:B\}$, which are assumed to be constant over the radar bandwidth and over the duration of $T_{\text{long}}$. If the time-varying radial distance of each $b^{\text{th}}$ point scatterer with respect to the radar is $r_b(t)$, then the approximate baseband received signal can be written as
\begin{equation}
\label{eq:RxSig}
\begin{split}
    y(\tau,t) \approx \sum_{b=1}^B a_b\, \text{rect} \left(\frac{\tau - \frac{2r_b(t)}{c}}{T_{\text{PRI}}} \right)\\
    \text{exp}\left(-j2\pi f_c \frac{2r_b(t)}{c}\right) \text{exp}\left(j\pi \gamma \left(\tau - \frac{2r_b(t)}{c}\right)^2\right),
    \end{split}
\end{equation}
where $c$ is the velocity of light. The model in (\ref{eq:RxSig}) is called the \emph{primitive model} or \emph{scattering center} model. The scattering center model is computationally simple to use to generate radar signatures, provided the positions and scattering coefficients of the scattering centers are available. Generally, the scattering centers are assumed to correspond to trackers placed on the live subject whose positions are gathered using MoCap technology. The position vector data ($\vec{r}_b$) of $B$ scatterers are spline interpolated from the video frame rate ($\frac{1}{T_f}$) to the pulse repetition frequency ($\frac{1}{T_{\text{PRI}}}$) of the radar. In prior works, the scattering coefficients or amplitudes, $a_b$, of the $B$ scatterers have been estimated from the size, shape and orientation of primitives of body parts corresponding to the scattering centers \cite{boulic1990global,mishra2019doppler}. For example, a marker placed on the human arm corresponds to an ellipsoid of dimensions comparable to the human arm. The approximate nature of the estimation of the scattering center coefficient results in very poor accuracy in the magnitude of the radar signatures. We propose to use the ray tracing results to obtain more accurate estimates of the scattering coefficients.

Based on the radar sampling frequency and pulse repetition frequency, the $n^{\text{th}}$ fast time sample of the discrete received signal for the $p^{\text{th}}$ PRI is
\begin{align}
\label{eq:DiscreteRxSig}
\begin{split}
    \boldsymbol{Y}[nT_s,pT_{\text{PRI}}] = \boldsymbol{Y}[n,p]=\sum_{b=1}^B a_b\, \text{rect} \left(\frac{n - \frac{2r_b[p]}{cT_s}}{N} \right)\\
    \text{exp}\left(-j2\pi f_c \frac{2r_b[p]}{c}\right) \text{exp}\left(j\pi \gamma \left(nT_s - \frac{2r_b[p]}{c}\right)^2 \right).
    \end{split}
\end{align}
Here $\{p = 1:LP\}$ correspond to the PRIs within $L$ CPIs while $\{n=1:N\}$ correspond to the fast time samples within one PRI as shown in the figure.
The ray tracing techniques provided the RCS estimates ($\sigma^{\text{vv}}[fT_f],\sigma^{\text{hh}}[fT_f]$) at $f_c$ for the whole human body at the video frame rate of the MoCap data. Since vertical co-polarization is the most commonly used framework in automotive scenarios, we interpolate $\sigma^{\text{vv}}[f], f= 1:F$ to radar pulse repetition frequency to get $\sigma^{\text{vv}}[p], p = 1:P$. The scattering coefficients can be assumed to be uniform across the radar bandwidth at automotive radar frequencies. Therefore, $\sqrt{\sigma^{\text{vv}}[p]}$ may be regarded as the first fast time sample of the scattered signal in (\ref{eq:DiscreteRxSig}), at  every $p^{\text{th}}$ PRI with $\gamma=0$ (ray tracing is applied at single frequency).
\begin{align}
\label{eq:LinFrame}
    \sqrt{\sigma^{\text{vv}}[p]} = \boldsymbol{Y}[n=1,p]=\sum_{b=1}^B a_b \, 
    \text{exp}\left(-j2\pi f_c \frac{2r_b[p]}{c} \right).
\end{align}
Now if we assume that for slow moving humans, the reflectivities of the point scatterers ($a_b$) fluctuate very slowly across $L$ CPIs ($T_{\text{long}}$) but the positions of the point scatterers ($\vec{r}_b$) change significantly across $M$ PRIs ($T_{\text{short}} = M \times T_{\text{PRI}}$), we can frame a linear regression model $\boldsymbol{\Phi} \boldsymbol{A} = \boldsymbol{\Psi}$ using
\begin{gather}
\label{eq:LinReg}
\boldsymbol{\Phi} = 
\begin{bmatrix}
    e^{-j2\pi f_c \frac{2 r_1[1]}{c}} & e^{-j2\pi f_c \frac{2 r_2[1]}{c}} & \dots &  e^{-j2\pi f_c \frac{2 r_B[1]}{c}}\\
    e^{-j2\pi f_c \frac{2 r_1[M]}{c}} & e^{-j2\pi f_c \frac{2 r_2[M]}{c}} & \dots &  e^{-j2\pi f_c \frac{2 r_B[M]}{c}}\\
    \vdots & \vdots  & \ddots & \vdots  \\
    e^{-j2\pi f_c \frac{2 r_1[LP]}{c}} & e^{-j2\pi f_c \frac{2 r_2[LP]}{c}} & \dots &  e^{-j2\pi f_c \frac{2 r_B[LP]}{c}}
\end{bmatrix}  
\end{gather}
and
\begin{gather}
    \boldsymbol{A} =
\begin{bmatrix}
    a_1   \\
    a_2 \\
    \vdots \\
    a_B
\end{bmatrix},
\boldsymbol{\Psi} = 
\begin{bmatrix}
    \sqrt{\sigma^{\text{vv}}[1]}   \\
    \sqrt{\sigma^{\text{vv}}[M]} \\
    \vdots \\
    \sqrt{\sigma^{\text{vv}}[LP]}
\end{bmatrix}.
\end{gather}
The integer number of rows of $\boldsymbol{\Phi} \in \mathbb{C}^{K \times B}$ is obtained by rounding $\lfloor\frac{LP}{M}\rfloor$ to the nearest integer. We estimate the reflectivities of the $B$ point scatterers by solving for $\boldsymbol{A}$ using ordinary least squares ($\min\limits_{\boldsymbol{A}} ||\boldsymbol{\Psi} - \boldsymbol{\Phi} \boldsymbol{A}||_2^2$) \cite{williams1990overdetermined}, as shown below
\begin{gather}
      \boldsymbol{A} = \left(\boldsymbol{\Phi}^T \boldsymbol{\Phi} \right)^{-1}\boldsymbol{\Phi}^T\boldsymbol{\Psi}.
\end{gather}
Once the scattering center coefficients are estimated, they can be used in (\ref{eq:DiscreteRxSig}) to obtain the radar received data $\boldsymbol{Y}[n,p]$.
The choices of $L$ (and thereby $T_{\text{long}}$) as well as $M$ (and $T_{\text{short}}$) are critical while $P$ is fixed by the radar specifications.  
Since humans are typically slow moving targets, low values of $T_{\text{short}}$ will result in very small changes between $r_b[p]$ and $r_b[p+M]$. This could result in singularity errors in the solution. On the other hand, large values of $T_{\text{short}}$ will result in long $T_{\text{long}}$ intervals which is undesirable since the scattering coefficients are unlikely to remain unchanged over long durations. 

In the above method, we have discussed how to estimate $a_b$ for a monostatic radar configuration of vertically polarized radar. However, the method can be easily modified to allow considerable flexibility in terms of radar carrier frequency, bandwidth, radar position, aspect angles and polarization. 
\begin{itemize}
    \item Depending on the polarization requirement of the simulation framework, we can generate radar data by selecting corresponding RCS values ($\sigma^{\text{vv}},\sigma^{\text{hh}},\sigma^{\text{vh}}$ and $\sigma^{\text{hv}}$) computed from ray tracing for $\boldsymbol{\Psi}$ in  (\ref{eq:LinReg}). 
    \item Similarly, we can change from monostatic to bistatic radar configuration in (\ref{eq:RxSig}). We can obtain the bistatic radar signatures by choosing the bistatic RCS values computed from ray tracing.
\end{itemize}
\begin{algorithm}[htbp]
\caption{Simulation of radar signatures for every $T_{\text{long}} = LP\,\,T_{\text{PRI}}$}
\label{alg:simulation_methodology}
     \textbf{Input:} MoCap data of $B$ point scatterers on the human body at video frame rate ($\frac{1}{T_f}$): $ \{\vec{r}_b[f T_f] = \vec{r}_b[f],\, b=1:B\}$
     \begin{enumerate}
    \item Implement ray tracing on three-dimensional poly-mesh structure obtained from stick figure for every frame $\sigma^{\text{vv}}[f], f = 1:F$
    \item Spline interpolate frames $\{f=1:F\}$ of position vector of point scatterer data along with the RCS derived from ray tracing from video frame rate to radar pulse repetition frequency ($\frac{1}{T_{\text{PRI}}}$) to obtain $\{p = 1: LP\}$ values.
    \begin{enumerate}
        \item[(i)] $\vec{r}_b[f]$ $\longrightarrow$ $\vec{r}_b[p]$
        \item[(ii)] $\sigma^{\text{vv}}[f]$ $\longrightarrow$ $\sigma^{\text{vv}}[p]$
    \end{enumerate}
    \item Formulate $\boldsymbol{\Phi} \in \mathbb{C}^{K \times B}$ where $K = \lfloor \frac{LP}{M} \rfloor \approx B$ such that
    \begin{equation*}
    \boldsymbol{\Phi} = 
\begin{bmatrix}
    e^{-j2\pi f_c \frac{2 r_1[1]}{c}} & \dots &  e^{-j2\pi f_c \frac{2 r_B[1]}{c}}\\
    e^{-j2\pi f_c \frac{2 r_1[M]}{c}} & \dots &  e^{-j2\pi f_c \frac{2 r_B[M]}{c}}\\
    \vdots  & \ddots & \vdots  \\
    e^{-j2\pi f_c \frac{2 r_1[LP]}{c}} & \dots &  e^{-j2\pi f_c \frac{2 r_B[LP]}{c}}
\end{bmatrix}. 
\end{equation*}
Also formulate
\begin{equation*}
\boldsymbol{A} =
\begin{bmatrix}
    a_1   \\
    a_2 \\
    \vdots \\
    a_B
\end{bmatrix},
\boldsymbol{\Psi} = 
\begin{bmatrix}
    \sqrt{\sigma^{\text{vv}}[1]}   \\
    \sqrt{\sigma^{\text{vv}}[M]} \\
    \vdots \\
    \sqrt{\sigma^{\text{vv}}[LP]}
\end{bmatrix}.
\end{equation*}
    \item Estimate the reflectivities of $B$ point scatterers by $\boldsymbol{A} = \left(\boldsymbol{\Phi}^T \boldsymbol{\Phi} \right)^{-1}\boldsymbol{\Phi}^T\boldsymbol{\Psi}$ using ordinary least squares minimization of $||\boldsymbol{\Psi} - \boldsymbol{\Phi} \boldsymbol{A}||_2^2$.
    \item \textbf{Output:} Model received radar signal using position and estimated reflectivities of point scatterers for every $n^{\text{th}}$ fast time sample of $p^{\text{th}}$ PRI.
    \begin{equation*}
    \begin{split}
    \boldsymbol{Y}[n,p] = \sum_{b=1}^B a_b\, \text{rect} \left(\frac{n - \frac{2r_b[p]}{cT_s}}{N} \right)
    \text{exp}\left(-j2\pi f_c \frac{2r_b[p]}{c} \right) \\\text{exp} \left( j\pi \gamma \left(nT_s - \frac{2r_b[p]}{c}\right)^2 \right).
    \end{split}
    \end{equation*}
    \item Use $\boldsymbol{Y}[n,p]$ to obtain three types of radar signatures:
    \begin{itemize}
    \item Implement 1D Fourier transform on $\boldsymbol{Y}[n,p]$ along fast time axis ($n$) for every $p^{\text{th}}$ PRI to obtain radar range-time signature ($\tilde{\boldsymbol{\chi}}^{\text{RT}}$).
    \item Implement 1D Fourier transform on $\boldsymbol{Y}[n,p]$ along slow time axis ($p$) for every CPI ($P$ PRIs) to obtain Doppler-time spectrogram ($\tilde{\boldsymbol{\chi}}^{\text{DT}}$).
    \item Implement 2D Fourier transform on $\boldsymbol{Y}[n,p]$ to obtain range-Doppler ambiguity plot ($\tilde{\boldsymbol{\chi}}^{\text{RD}}$).
    \end{itemize}
    \end{enumerate}
\end{algorithm}
\subsection{Generation of radar signatures}
The two-dimensional radar data $\boldsymbol{Y}[n,p]$ along the fast and slow time axes are processed through Fourier transform to obtain three types of radar signatures for every $T_{\text{long}}$ duration. The three signatures are: range-time ($\tilde{\boldsymbol{\chi}}^{\text{RT}}$), Doppler-time ($\tilde{\boldsymbol{\chi}}^{\text{DT}}$) and time-varying range-Doppler ambiguity plots ($\tilde{\boldsymbol{\chi}}^{\text{RD}}$). 
As mentioned earlier, each $T_{\text{long}}$ interval of the radar data consists of $L$ CPIs, where each CPI is of $P$ PRIs. 

The \emph{range-time profile} is generated by implementing the one-dimensional Fourier transform on $\boldsymbol{Y}[n,p]$ along the fast time axis for each $p^{\text{th}}$ PRI as shown in
\begin{gather}
\label{eq:HRRP}
    \begin{split}
    \tilde{\boldsymbol{\chi}}_p^{\text{RT}}[g\Delta r] = \tilde{\boldsymbol{\chi}}_p^{\text{RT}}[g] = \sum_{n=1}^{N} \boldsymbol{Y}[n,p] \boldsymbol{H}_{1D}[n]e^{-j\frac{2\pi gn}{N}},\\
    g = \frac{-N}{2}:\frac{N}{2}-1
    \end{split}
\end{gather}
where, $\Delta r=\frac{c}{2 BW}$ is range resolution and $\boldsymbol{H}_{1D}[\cdot] \in \mathbb{R}^{N \times 1}$ is a one-dimensional window function.

The \emph{Doppler velocity spectrogram} is generated by implementing the one-dimensional Fourier transform on $\boldsymbol{Y}[n=1,p]$ along the slow time axis for each $l^{\text{th}}$ CPI as shown in
\begin{gather}
\label{eq:STFT}
    \begin{split}
    \tilde{\boldsymbol{\chi}}_l^{\text{DT}}[d\Delta f_D] =\tilde{\boldsymbol{\chi}}_l^{\text{DT}}[d] 
    = \sum_{p=(l-1)P+1}^{lP} \boldsymbol{Y}[n=1,p] \boldsymbol{H}_{1D}[p]e^{-j\frac{2\pi dp}{P}},\\
    d = \frac{-P}{2}:\frac{P}{2}-1
    \end{split}
\end{gather}
where, $\Delta f_D = \frac{1}{PT_{\text{PRI}}}$ is Doppler resolution.

\emph{Range-Doppler ambiguity plots} are generated for each $l^{\text{th}}$ CPI through two-dimensional Fourier transform of $\boldsymbol{Y}[n,p]$ along the fast and slow time axes as shown below
\begin{align}
    \label{eq:2Dfft}
     \tilde{\boldsymbol{\chi}}_l^{\text{RD}}[g,d] =
    \sum_{p=(l-1)P+1}^{lP} \sum_{n=1}^{N} \boldsymbol{Y}[n,p] \boldsymbol{H}_{2D}[n,p] e^{-j\frac{2\pi gn}{N} }e^{-j\frac{2\pi dp}{P}},
\end{align}
where, $\boldsymbol{H}_{2D}[\cdot] \in \mathbb{R}^{N \times P}$ is a two-dimensional window function. The process is repeated across all the $L$ CPIs to obtain the time-varying range-Doppler ambiguity plots.

Algorithm \ref{alg:simulation_methodology} summarizes the proposed simulation methodology to generate the radar signatures with accurate scattering center coefficients for every $T_{\text{long}}$ period.
\section{Experimental Results}
In this section we present the experimental results for validating the proposed methodology. We collect MoCap data of human motion and use it to simulate radar data. Simultaneously, we collect hardware based radar data for the same human subject from a measurement setup at 77 GHz. We perform ray tracing on the three-dimensional poly-mesh structure obtained from the stick figure animation of every frame of the MoCap data to simulate the monostatic RCS of the pedestrian for vertical polarization. Then we use these values to generate scattering coefficients of the scattering center model of a human. Finally, we generate the simulated radar signatures which we compare with measurement results. 
\subsection{Experimental data collection}
\label{Sec:ExpSetUp}
\begin{figure}[!ht]
    \centering
    \includegraphics[width = 0.45\textwidth]{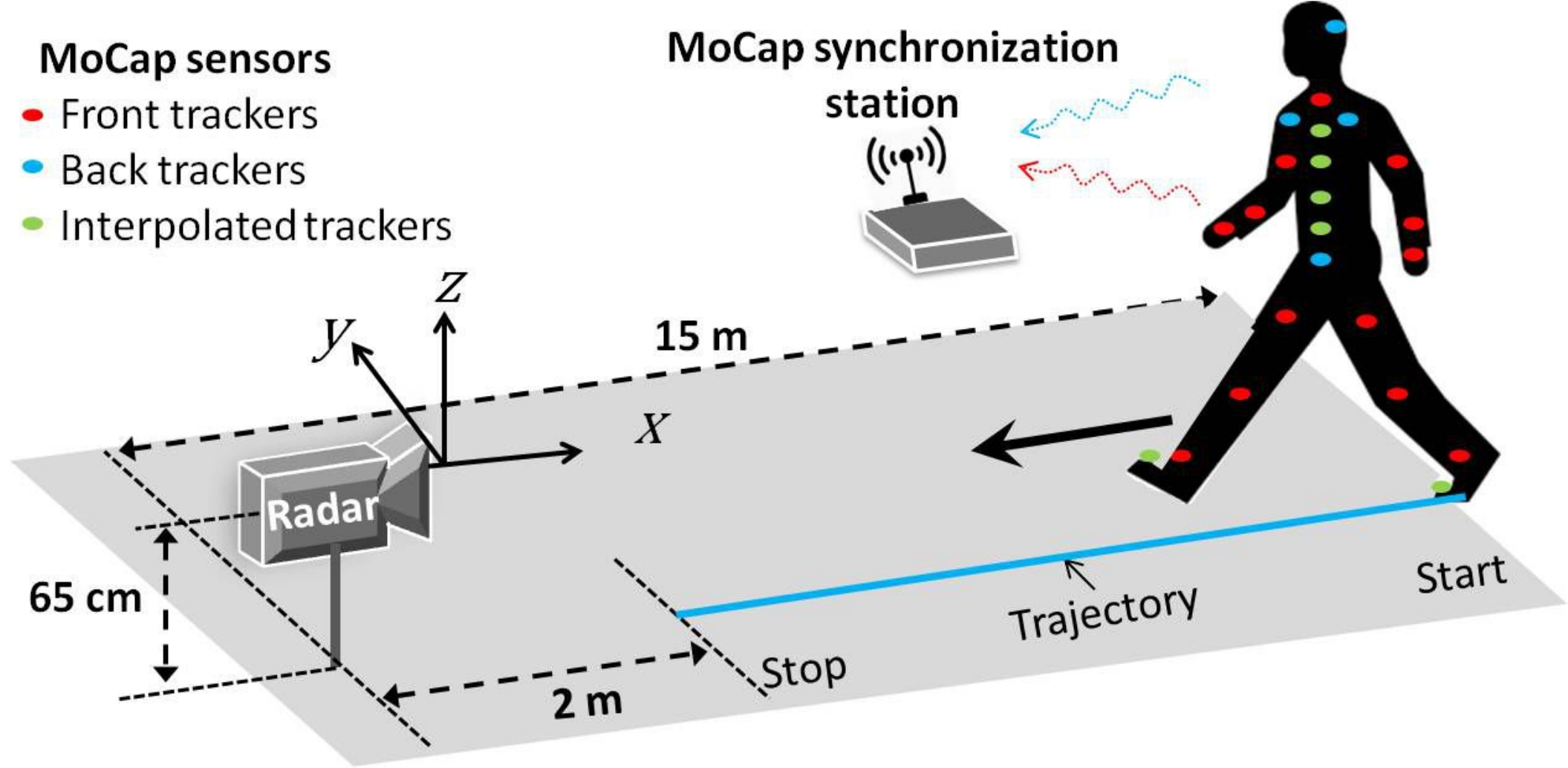}
    \caption{Subject wearing 17 trackers in front (red) and back (blue) view; 6 interpolated trackers (green), for collecting MoCap data, walks radially towards the radar sensor (INSAR RadarLog) from 15 m distance and stops 2 m before the radar along the trajectory (indicated by blue line). The radar sensor is positioned at [0, 0, 0.65] m. The wireless communication between the MoCap sensors and the synchronization station takes place at 60\,Hz frame rate.}
    \label{fig:radar_scenario}
\end{figure}

We present the experimental setup in this section.
We consider a human subject moving along the trajectory shown in Fig.~\ref{fig:radar_scenario}.
We collect MoCap data of the human motion using Xsens MTw Awinda \cite{xsens}, an inertial measurement unit containing three-dimensional linear accelerometers and rate gyroscopes. 17 trackers (front and back side) are attached to defined locations on the subject's body to measure the motion of each body segment. Additionally, position information of 6 other body segments on the torso and feet are determined by interpolation by the MoCap software. Wireless communication between the sensors and the synchronization station takes place at 60\,Hz frame rate. The MoCap data of the 23 markers are used for simulating the radar returns. For validation purposes, the radar returns from the subject is simultaneously captured using a 77 GHz linear frequency modulated INRAS RadarLog sensor \cite{inrasradarlog,held2018radar}.
\begin{table}[h]
    \centering
    \begin{tabular}{cc}
    \hline
    \hline
    \multicolumn{2}{c}{Radar parameters}\\
    \hline
    \hline
     Carrier frequency ($f_c$) & 77 GHz \\
     Bandwidth ($BW$) & 2 GHz\\
     Sampling frequency ($f_s$) & 10 MHz\\
     Up Chirp duration ($T_{\text{upchirp}}$) & 51.2 $\mu$s\\
     Pulse repetition interval ($T_{\text{PRI}}$) & 61.2 $\mu$s\\
     No. of chirps per CPI ($P$) & 1024 \\
     Range resolution ($\Delta r$) & 7.5 cm\\
     Doppler resolution ($\Delta f_D$) &  15.9 Hz\\
     Radar sensor position & [0, 0, 0.65] m\\
    \hline 
    \hline\\
\end{tabular}
\caption{Radar parameters used for simulation are chosen to match the INRAS RadarLog sensor. }
\label{tab:RadarParam}
\end{table}
The simulation parameters for the radar signal model discussed in section \ref{Sec:SimRadarModel} are chosen to match the radar hardware configurations as listed in Table \ref{tab:RadarParam}.
\subsection{Results from electromagnetic ray tracing}
\label{Sec:RayTracingResults}
\begin{figure*}[!ht]
    \centering
    \subfloat[\label{fig:Monostatic_front_77GHz}]{\includegraphics[width=0.3\textwidth]{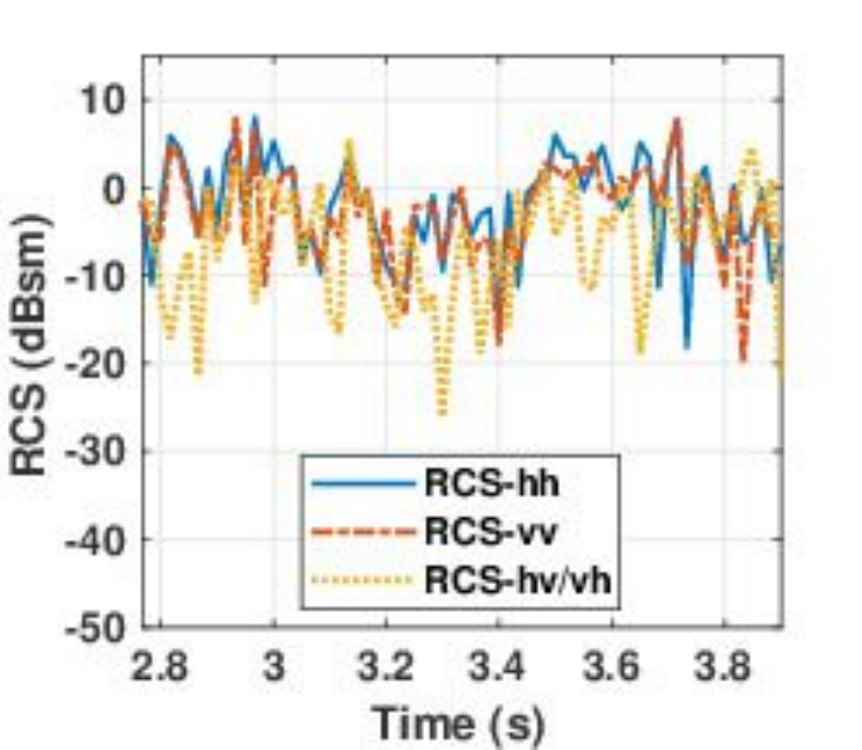}}\hfil
    \subfloat[\label{fig:Monostatic_45deg_77GHz}]{\includegraphics[width=0.3\textwidth]{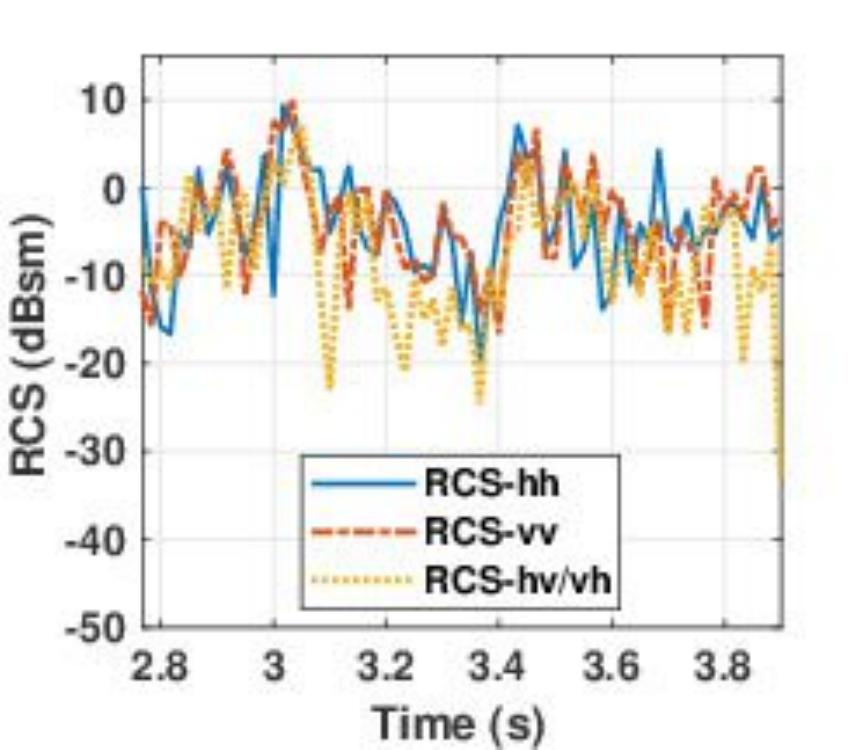}}\hfil
    \subfloat[\label{fig:Monostatic_90deg_77GHz}]{\includegraphics[width=0.3\textwidth]{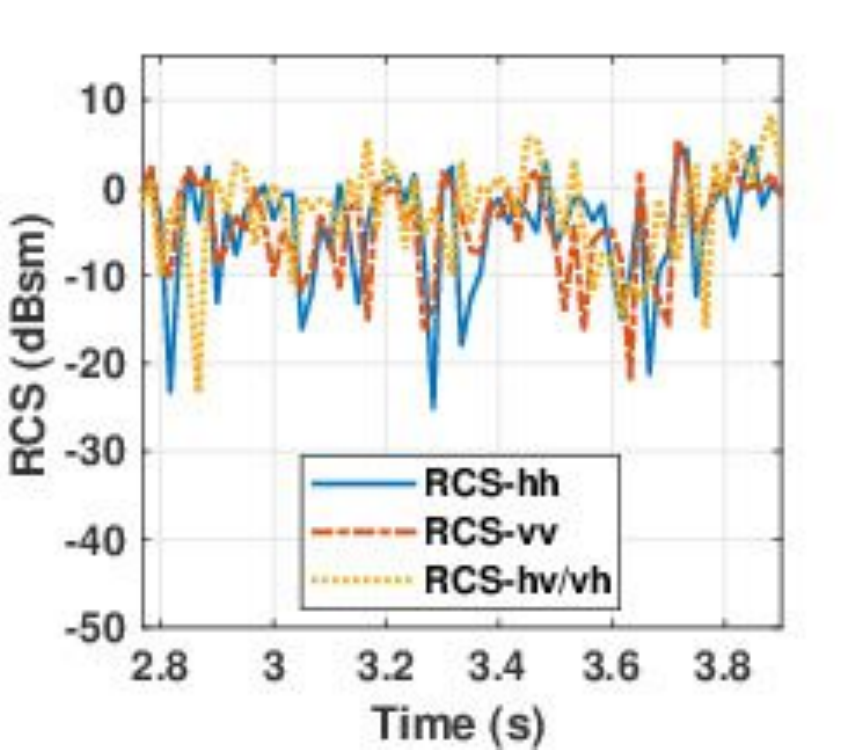}}
    \caption{Simulated monostatic ($\phi^{i} = \phi^s $) RCS across multiple frames corresponding to one walking stride obtained from ray tracing at 77 GHz for three aspect angles (a) front incidence ($\phi^{i} = 0^{\circ}$) (b) oblique incidence ($\phi^{i} = 45^{\circ}$) and (c) $90^{\circ}$ incidence ($\phi^{i} = 90^{\circ}$).}
    \label{fig:Monostatic}
\end{figure*}

The animated stick figure model obtained from MoCap is embodied using an in-built library of a nude male in Poser Pro software from Smith Micro Inc.\cite{poser}. Each frame of the human body is subsequently rendered into a three-dimensional poly-mesh structure composed of 3052 triangular facets. The data for each frame consist of three-dimensional position coordinates of the triangle's vertices which are exported to MATLAB for further processing.
The simulations are performed for both co-polarization and cross-polarization scenarios. Based on the scattered signal from all the body parts, we estimate the total monostatic RCS of the human at every frame of the MoCap data. We present the results for a complete walking stride - the full swing motion of a hand/leg - of 69 frames from $2.8\,$s to $3.9\,$s in Fig.~\ref{fig:Monostatic} for different incident aspect angles. 
The figure shows that the co-polarization ($\sigma^{\text{vv}}$ and $\sigma^{\text{hh}}$) components range from -10\,dBsm to +5\,dBsm. The cross-polarization components are generally weaker by approximately 10\,dB. These RCS values are within the range of those reported from measurement studies of pedestrians at X-band frequencies \cite{yamada2005radar, le2007numerical}. 
The versatility of the ray tracing methodology allows us to compute RCS at a variety of radar configurations including carrier frequencies, polarizations and radar positions. In the Appendix A, we present the monostatic radar RCS at 24 GHz, the other popular automotive radar frequency. We also provide the bistatic RCS values at both 24 and 77 GHz.

All the results presented in this section were generated with ray tracing alone. The next set of results are generated by hybridization of ray tracing and point scatterer modeling. We use the vertical co-polarized RCS ($\sigma^{\text{vv}}$) values at front incidence ($\phi^{i}=0^{\circ}$) to match the radar hardware configurations.
\subsection{Discussion on parameters}
\begin{figure}[!ht]
    \centering
     \subfloat[\label{fig:Avg_est_error_vs_K}]{\includegraphics[width = 0.23\textwidth,height = 1.5in]{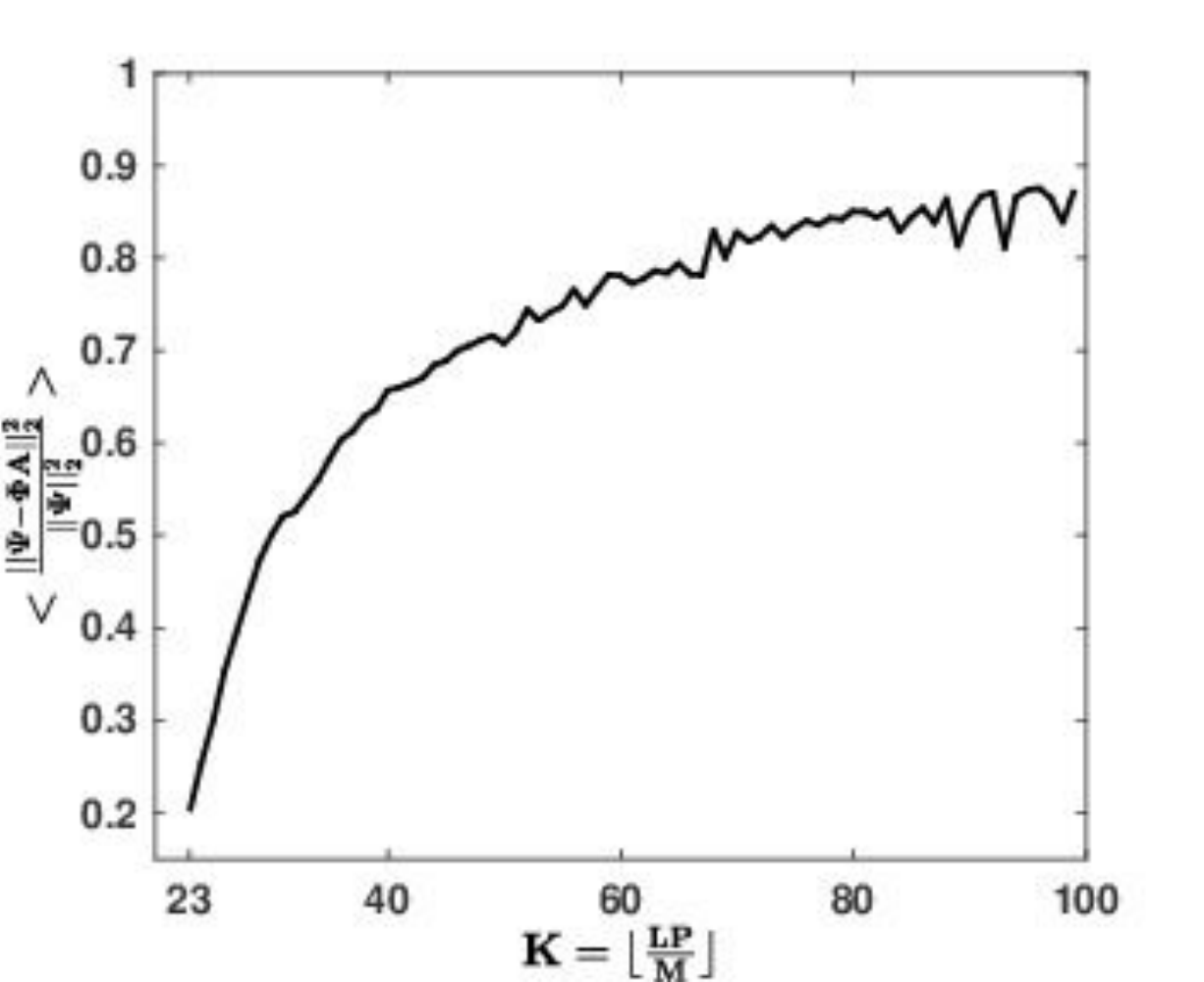}}
     \subfloat[\label{fig:NMSE_vs_M}]{\includegraphics[width=0.23\textwidth,height = 1.5in]{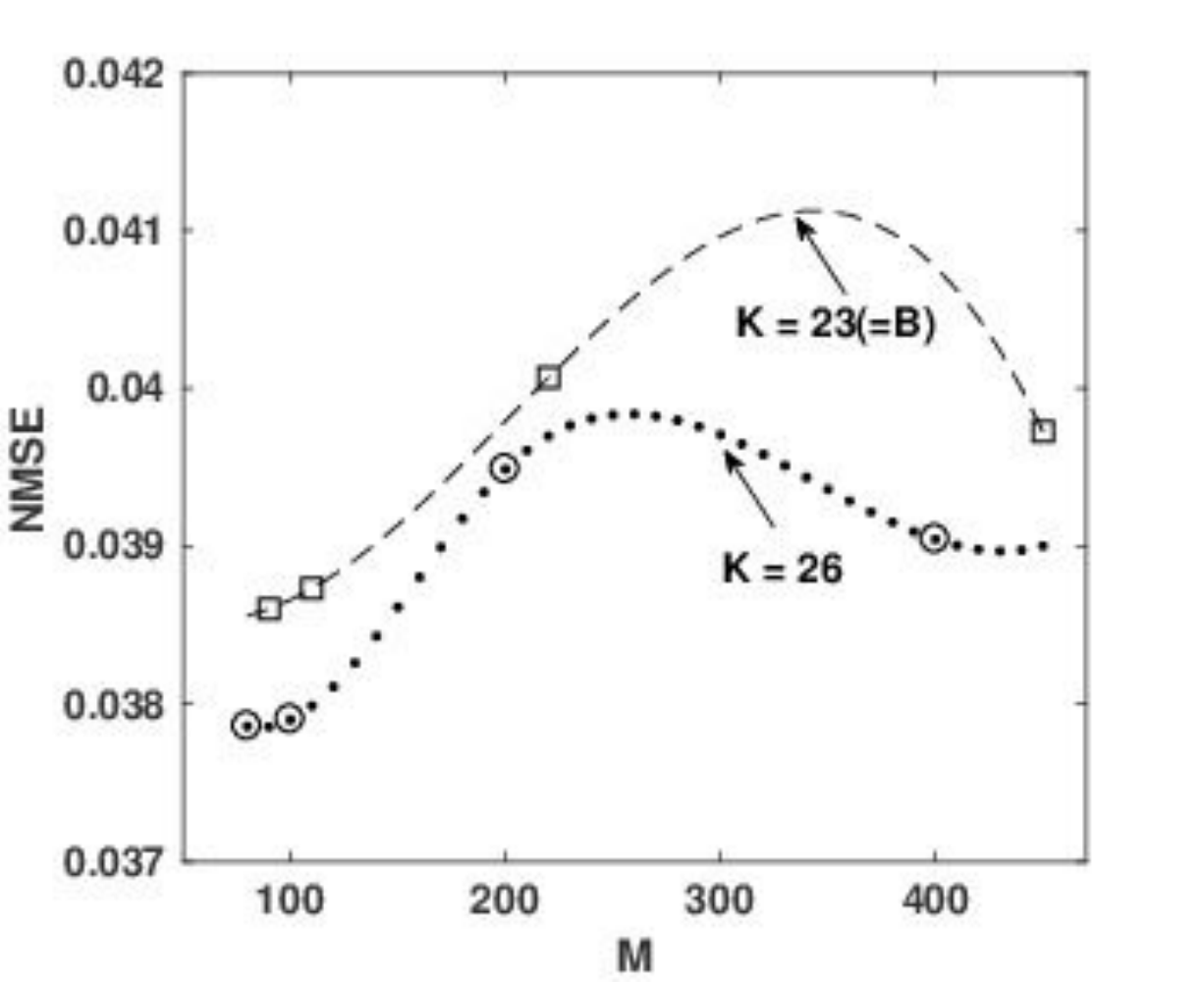}}
    \caption{To make the choice of $T_{\text{long}}$ (and $L$) and $T_{\text{short}}$ (and $M$): (a) Estimation error $\frac{||\boldsymbol{\Psi} - \boldsymbol{\Phi} \boldsymbol{A}||_2^2}{||\boldsymbol{\Psi}||_2^2}$ averaged over the number of estimations done for duration of target motion for different values of $K = \lfloor \frac{LP}{M}\rfloor$ is plotted. (b) Comparing the NMSE between simulated and measured range-time plots for $K = 23$ and $K = 26$ for different values of $M$.}
    \label{fig:Avg_est_error_plot}
\end{figure}

The scattering coefficients are estimated by solving the linear regression framework in (\ref{eq:LinFrame}) where $T_{\text{long}}$ (and $L$) and $T_{\text{short}}$ (and $M$) have to be carefully chosen. Both $L$ and $M$ determine $K$, the number of rows in $\boldsymbol{\Phi}$ matrix, since $K$ is rounded to the nearest integer $\lfloor\frac{LP}{M}\rfloor$ and $P$ is fixed. 
Fig.~\ref{fig:Avg_est_error_vs_K} shows the average $l_2$ norm error, $\frac{||\boldsymbol{\Psi} - \boldsymbol{\Phi} \boldsymbol{A}||_2^2}{||\boldsymbol{\Psi}||_2^2}$ for different values of $K$. When $K$ is very large due to small values of $M$, we get very high errors. This is because for slow moving targets, such as humans, there is very small variation in the position of some of the scatterers (such as torso) in consecutive $T_{\text{short}}$ intervals. This results in singularities in the problem formulation. We find that the optimum results occur when $K \approx B$, that is when the $\boldsymbol{\Phi}$ matrix is close to a square matrix.
Different combinations of $L$ and $M$ can result in similar values of $K$. However, when $M$ is very large, this gives rise to a correspondingly large value of $L = \frac{MK}{P}$. But long $T_{\text{long}}$ duration is undesirable since scattering coefficients are likely to fluctuate over long intervals due to variations in target aspect. We compared the NMSE of the measured and simulated range-time ambiguity plots for different $M$ for a fixed $K = 23(=B)$ and $K = 26$ in Fig.~\ref{fig:NMSE_vs_M}. The result shows that the NMSE is lowest for slightly over-determined matrix when $K = 26(= B+3)$.
Based on the above studies, we determined $M =80$ and $L = 2$ to be the optimum values for our simulation.
\subsection{Radar signatures generated from simulated and measured radar data}
\label{Sec:ResultsSig}
Based on the choice of $M$ and $L$, the $T_{\text{long}}$ and $T_{\text{short}}$ used in the linear regression framework are $12.5\,$ms and $4.9\,$ms respectively. 
We present three types of radar signatures - the high range-resolution profile, the Doppler-time spectrogram and the range-Doppler ambiguity plots and compare these signatures with those generated from measurement data collected from the radar hardware. The measurement data is suitably range compensated to obtain the time-varying radar cross-section of the target. Since the measurement data is naturally corrupted by noise, an ordered statistics constant false alarm rate (OS-CFAR) algorithm based on \cite{ludloff2013praxiswissen} is implemented on the measurement data, which adaptively estimates the detection threshold for each cell based on neighboring cells. The CFAR algorithm is not required on the simulation data where noise is not considered. We present both qualitative and quantitative comparisons between the simulated and measured radar signatures. 

\begin{figure*}[!ht]
\centering
    \subfloat[\label{Subfig:RTresultsA}]{\includegraphics[width=0.3\textwidth]{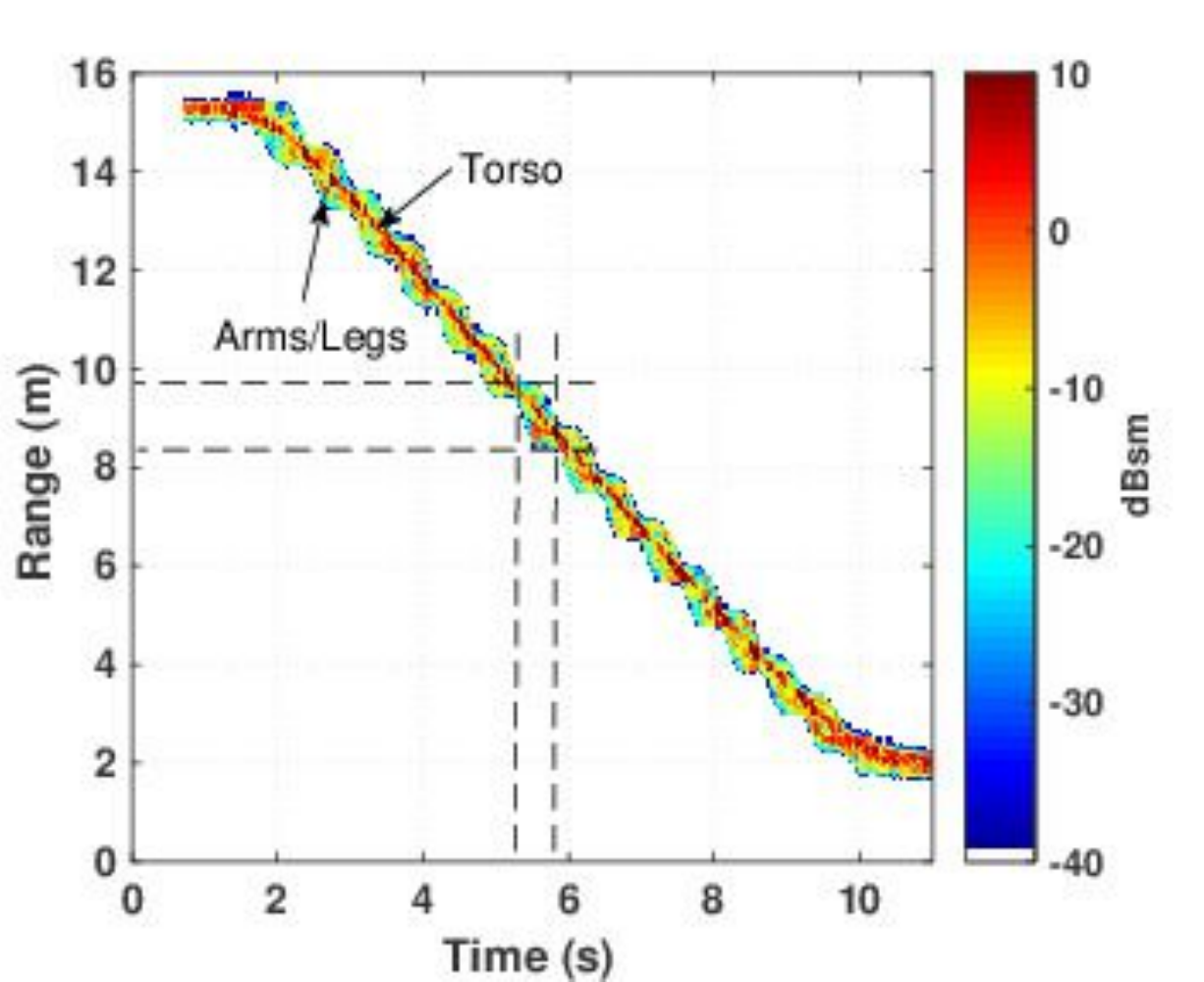}}
    \subfloat[\label{Subfig:DTresultsA}]{\includegraphics[width=0.3\textwidth]{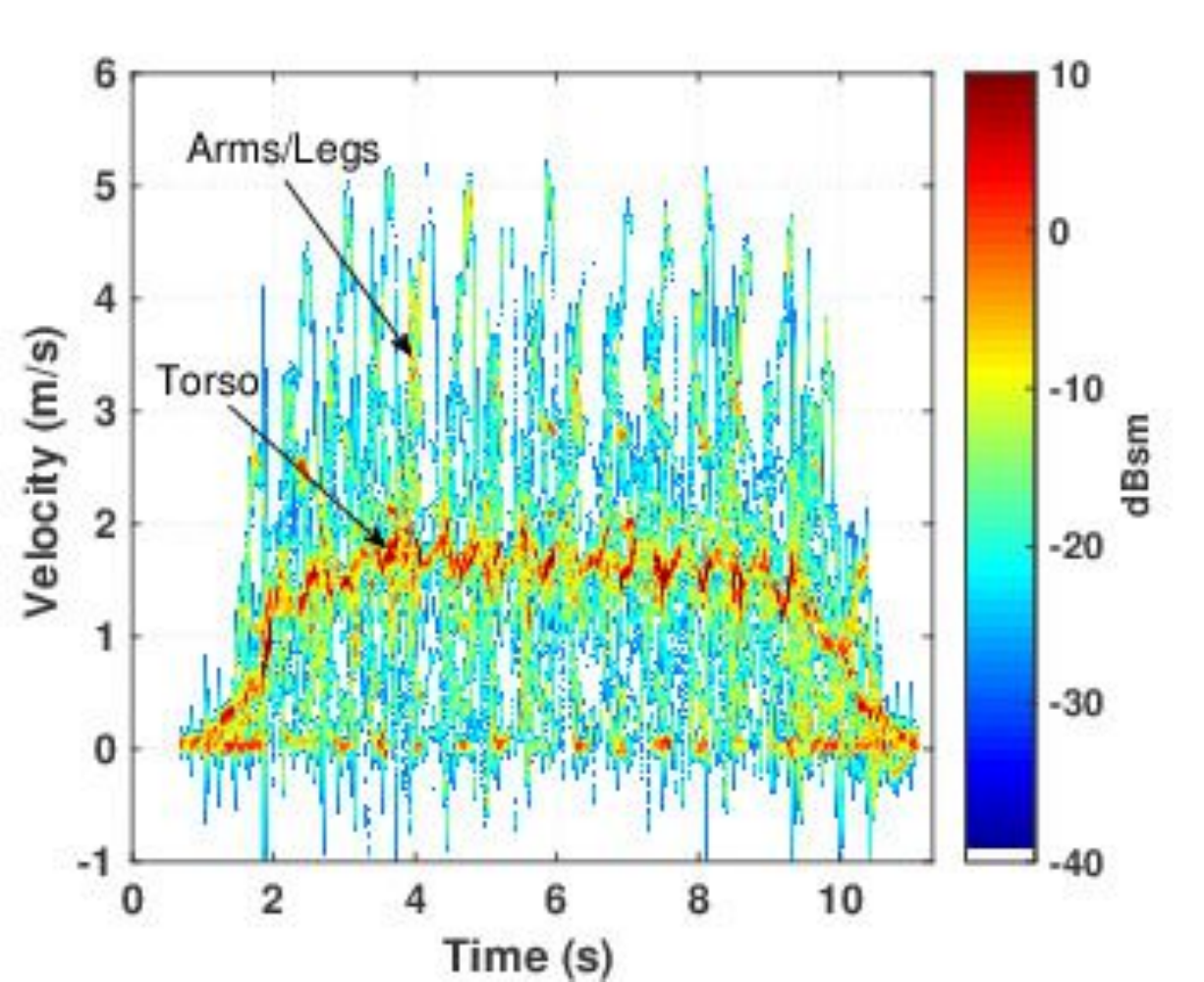}}
    \subfloat[\label{Subfig:RDresultsA}]{\includegraphics[width=0.3\textwidth]{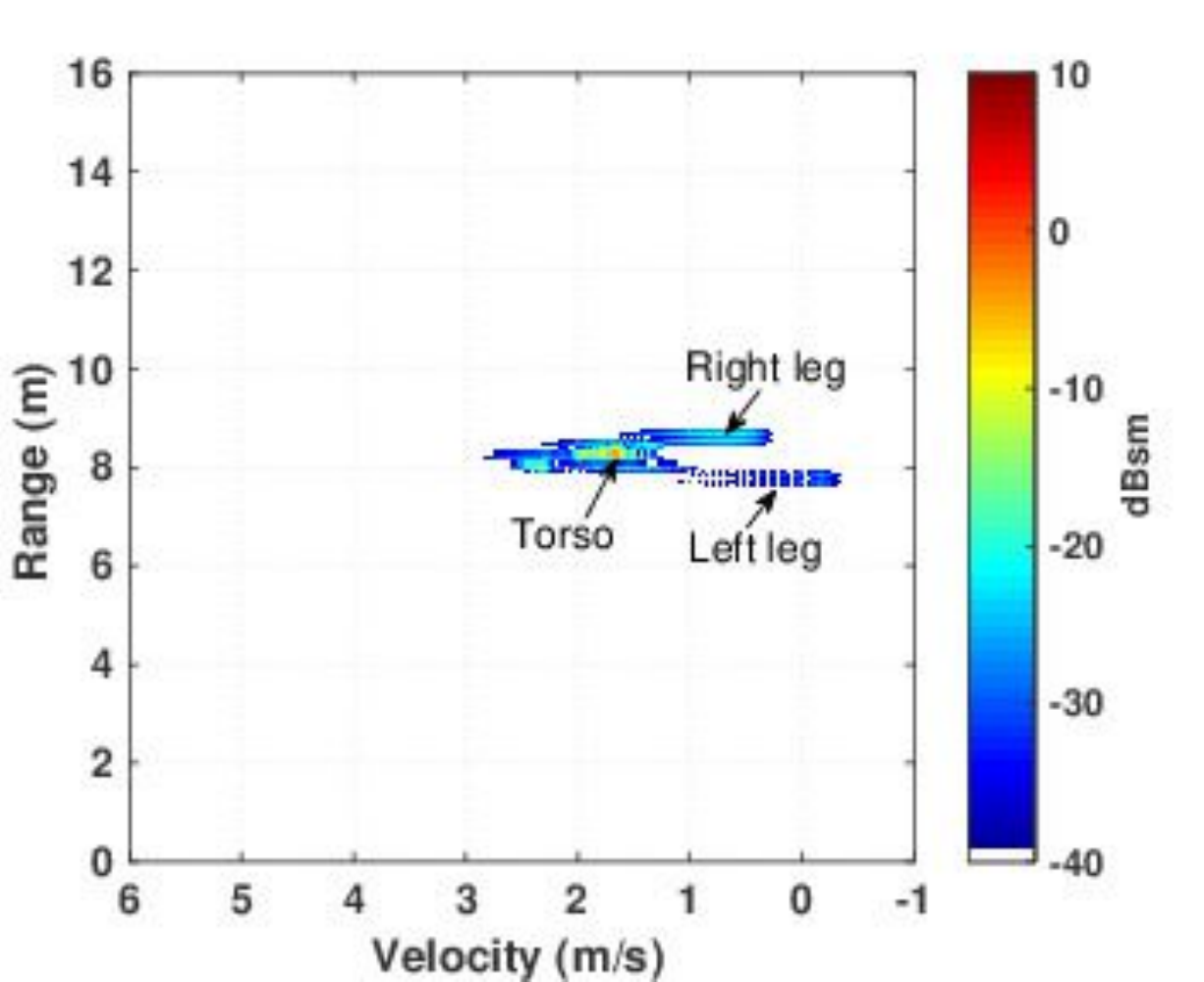}}
    \caption{Simulated radar signatures of a human walking towards a 77 GHz monostatic radar. (a) range-time ambiguity plot; (b) Doppler-time ambiguity plot; and (c) range-Doppler ambiguity plot for one CPI (from 6.16 to 6.22 seconds).}
\end{figure*}

\begin{figure*}[!ht]
\centering
    \subfloat[\label{Subfig:RTresultsB}]{\includegraphics[width=0.3\textwidth]{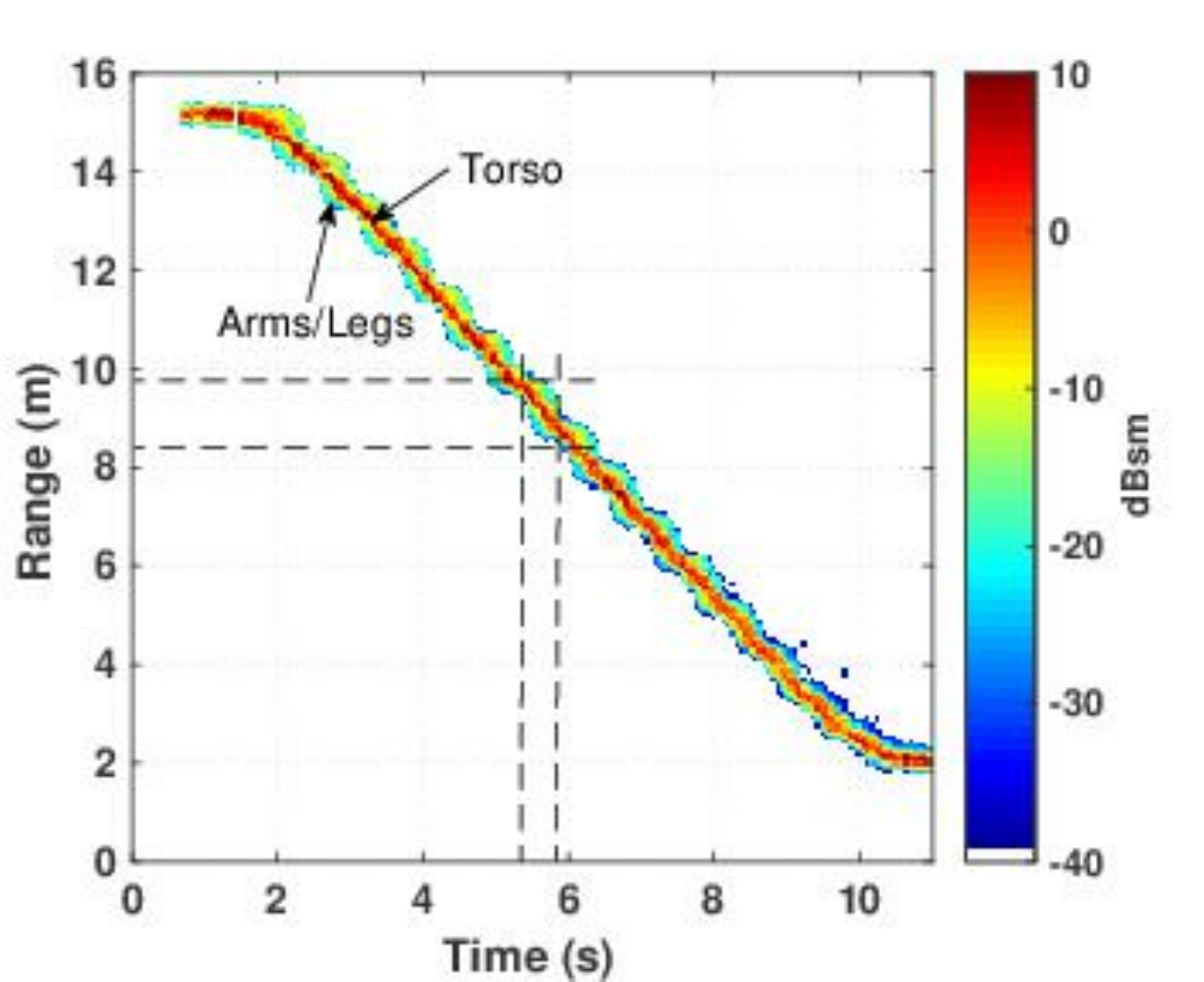}}
    \subfloat[\label{Subfig:DTresultsB}]{\includegraphics[width=0.3\textwidth]{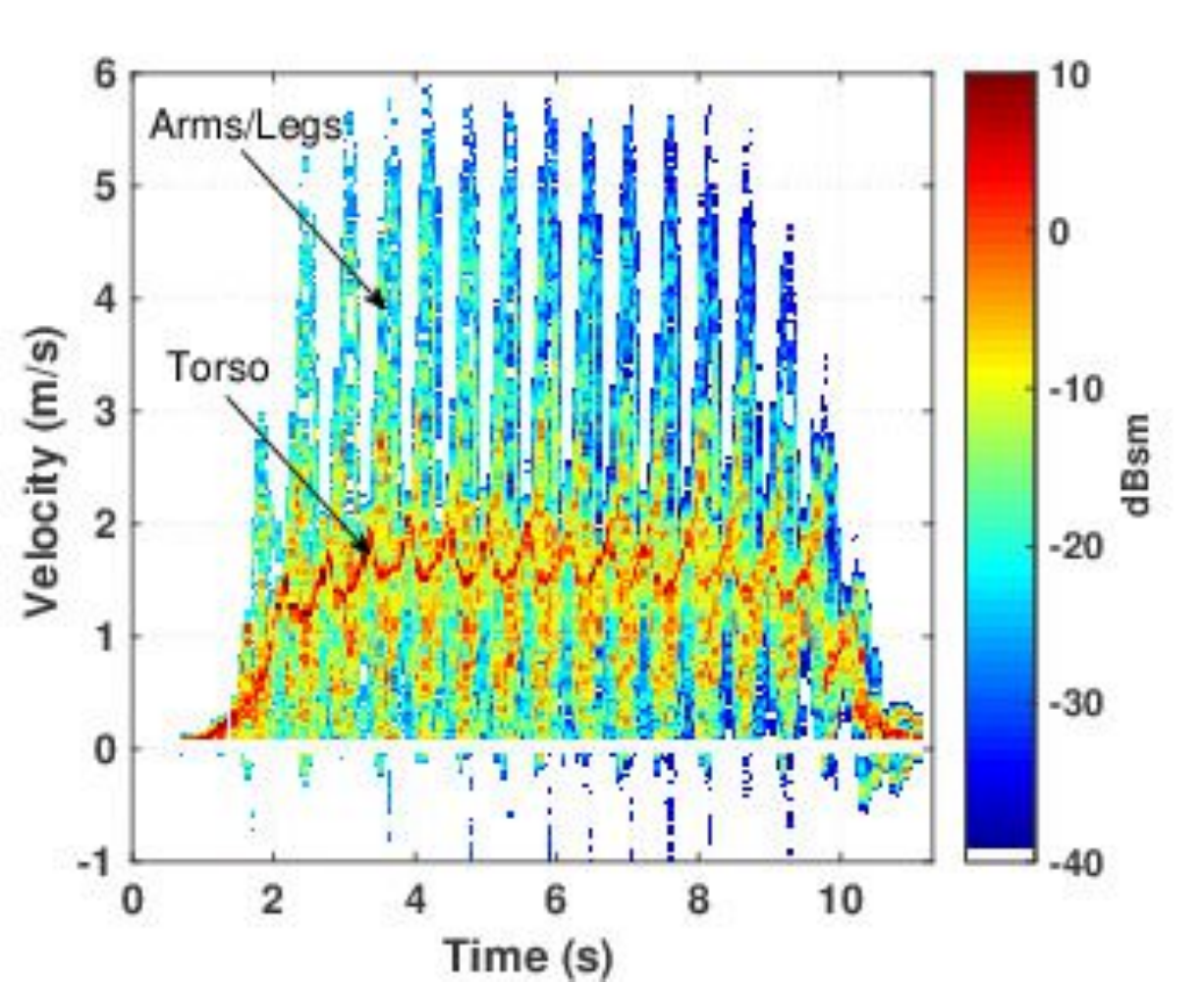}} \subfloat[\label{Subfig:RDresultsB}]{\includegraphics[width=0.3\textwidth]{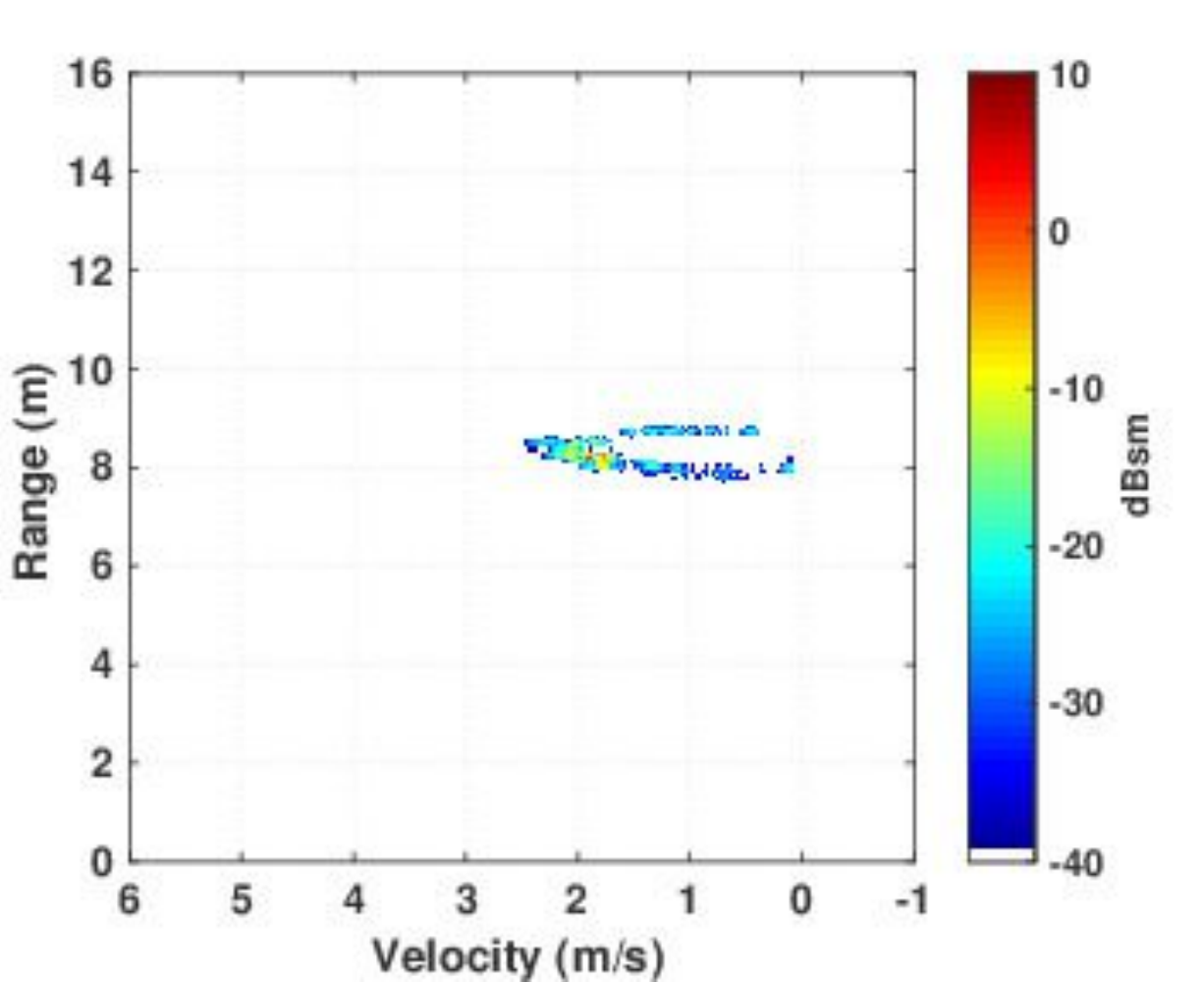}}
    \caption{Measured radar signatures of a human walking towards a 77 GHz monostatic radar. (a) range-time ambiguity plot; (b) Doppler-time ambiguity plot; and (c) range-Doppler ambiguity plot for one CPI (from 6.16 to 6.22 seconds).}
\end{figure*}

First, we present the \emph{high range-resolution profile} of the walking human in Fig.~\ref{Subfig:RTresultsA} and Fig.~\ref{Subfig:RTresultsB}. The figure on the top is generated from simulation data ($\tilde{\boldsymbol{\chi}}^{\text{RT}}$) while the one in the bottom is from measurement data ($\boldsymbol{\chi}^{\text{RT}}$). Values below -40 dBsm threshold are not shown in both the figures. We observe that the human is first stationary for 1.5 s and then approaches the radar from a range of 15 m to 2 m from 1.5 s to 10.3 s. The swinging motion of the arms and legs give rise to micro-range features about the torso that spans approximately 1.5 m. The range ambiguity is 7 cm. Therefore, it is difficult in both figures to resolve the independent point scatterers from the different body parts along range. The simulated results closely resembles the measured results in terms of dynamic range. The torso appears to be the strongest component in both images when compared to the arms and legs. The range spread due to the spatial extent of the target is nearly identical in both the images (indicated by horizontal dashed lines). The vertical dashed lines in both the figures indicate the similarity in time span also. Thus visually, there is structural similarity in the images. \\
Next, we examine the \emph{Doppler-spectrograms} from the simulated ($\tilde{\boldsymbol{\chi}}^{\text{DT}}$) and measured data ($\boldsymbol{\chi}^{\text{DT}}$) in Fig.~\ref{Subfig:DTresultsA} and Fig.~\ref{Subfig:DTresultsB}. Since the human is approaching the radar, the Dopplers are mostly positive with some negative Dopplers due to the back swing of the arms and legs. The human is walking at a velocity of approximately 1.5 m/s. This results in a strong torso Doppler component in both images. We can observe much weaker micro-Dopplers from the arms and legs up to velocities of 5 m/s. The Doppler span for measurement results are slightly higher than the simulated results due to noise characteristics. The periodicity of the strides in the two figures shows excellent agreement. There is a strong DC component in the simulation figure that is not present in the measurement results due to a DC filter in the radar hardware to eliminate static clutter.\\
Finally, we present the \emph{range-Doppler ambiguity plot} for a single CPI (from 6.16 to 6.22 seconds) in Fig.~\ref{Subfig:RDresultsA} and Fig.~\ref{Subfig:RDresultsB}. Here we observe that the range and Doppler ambiguities in both the simulated ($\tilde{\boldsymbol{\chi}}^{\text{RD}}$) and measured data ($\boldsymbol{\chi}^{\text{RD}}$) are nearly identical. We are now able to resolve the arms, legs and torso in the ambiguity plots. The simulation result enables us to correctly identify the different body parts. Again the peak and dynamic range of the two plots are very similar.

\begin{figure}[!ht]
    \centering
    {\includegraphics[width=0.35\textwidth]{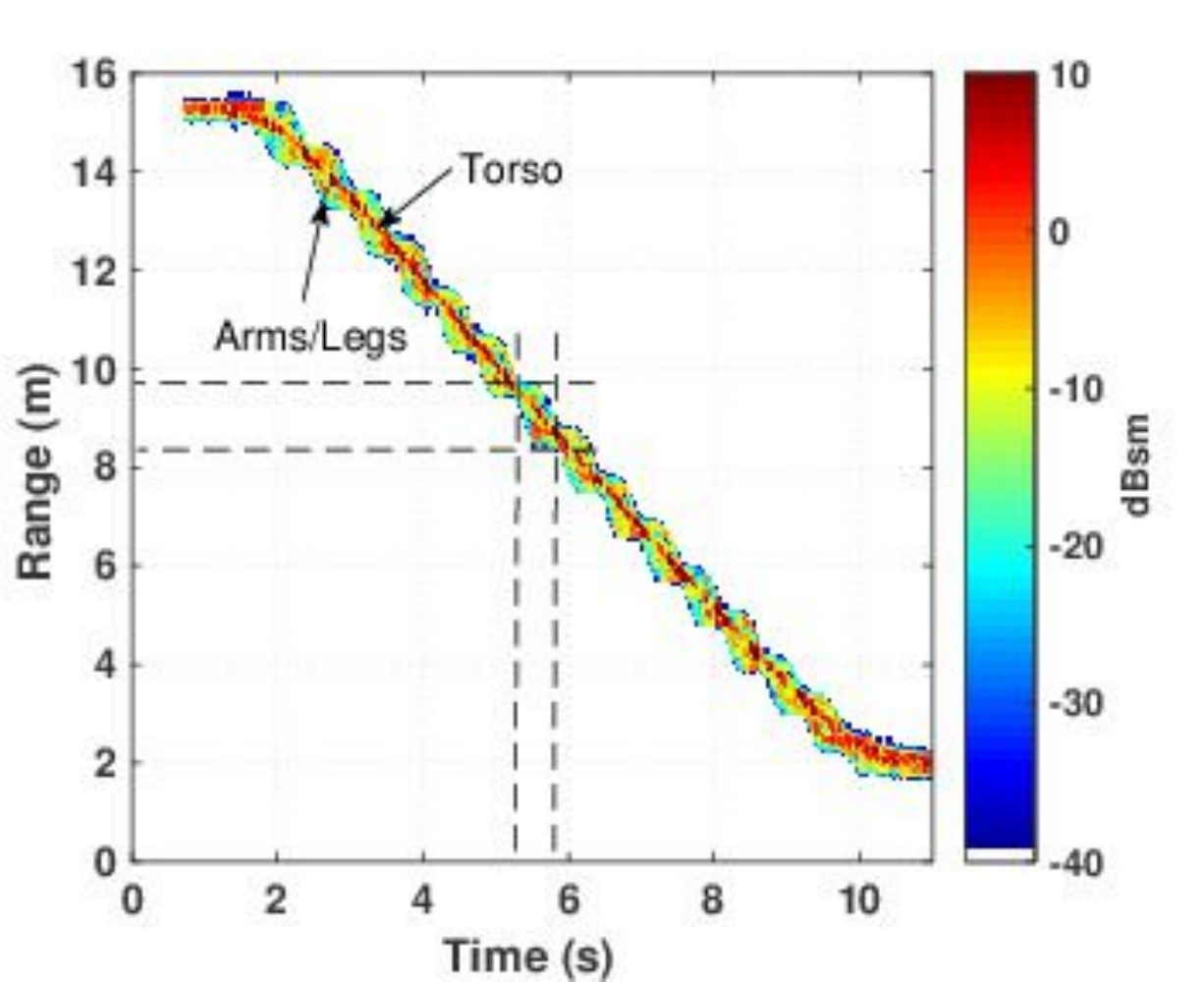}}
    \caption{SSIM and NMSE values for range-Doppler plots for one complete walking stride from 5.76 to 6.89 seconds. }
    \label{fig:SSIM_NMSE}
\end{figure}

In the above discussion, we have qualitatively compared the simulated and measured results. Next, we perform a quantitative comparison between the two signatures in the form of two metrics - the normalized mean square error (NMSE) and the structural symmetry index (SSIM). The NMSE for the range time plot is computed by
\begin{gather}
\label{eq:NMSE}
     NMSE = \frac{||\tilde{\boldsymbol{\chi}}^{\text{RT}}-\boldsymbol{\chi}^{\text{RT}}||^2_2}{||\boldsymbol{\chi}^{\text{RT}}||^2_2}.
\end{gather}
The SSIM is a metric used for comparing structural differences such as luminance and contrast between two images \cite{wang2004image}. It is computed by
\begin{gather}
\label{eqn:SSIM_Simple}
SSIM = \frac{(2E[\tilde{\boldsymbol{\chi}}]E[\boldsymbol{\chi}])(2\text{covar}[\tilde{\boldsymbol{\chi}},\boldsymbol{\chi}])}{(E^2[\tilde{\boldsymbol{\chi}}]+E^2[\boldsymbol{\chi}])(\text{var}[\tilde{\boldsymbol{\chi}}]+\text{var}[\boldsymbol{\chi}])},
\end{gather}
where $E[\cdot]$, $\text{var}[\cdot]$ and $\text{covar}[\cdot]$ denote mean, variance and co-variance of the two images. When the images are identical, its value is 1.
\begin{table}[htbp]
    \centering
    \begin{tabular}{cccc}
    \hline
    \hline
      & Range-time &  Doppler-spectrograms & Range-Doppler\\
      \hline
      \hline
      SSIM & 0.86 & 0.81 & 0.99 \\
      NMSE & 0.04 & 0.10 & 0.03 \\
      \hline
      \hline\\
    \end{tabular}
    \caption{Quantitative comparison between simulated and measured range-time, Doppler-time and range-Doppler plots through NMSE and SSIM values for the duration of target motion}
    \label{tab:All_SSIM_NMSE}
\end{table}
Table \ref{tab:All_SSIM_NMSE} shows the NMSE and SSIM for the three radar signatures for the duration of the target motion. All three signatures show low values of NMSE, and SSIM values close to 1 which indicates the close similarity between the simulation and measurement data. Fig.~\ref{fig:SSIM_NMSE} shows the NMSE and the SSIM between the simulated and measured range-Doppler ambiguity plots over the duration of one walking stride (9 $T_{\text{long}}$) from 5.76 to 6.89 seconds. 
The results in the figure show the range of SSIM between 0.96 to 0.99 which is close to ideal. The NMSE is likewise close to zero. 

The computational complexity of the proposed approach required ray tracing to be carried out at video frame rate, matrix inversion operations for determining scattering coefficients and linear operations for point scatterer modeling at radar sampling frequencies. The matrix inversion operation is computationally not very hard due to the small size of the matrix ($[(K \approx B) \times B]$). Among these three steps, the ray tracing operation is the most computationally expensive and hence we discuss its complexity in the Appendix.
\section{Conclusion}
\label{Sec:Conclusion}
The shooting and bouncing ray technique based on ray tracing and geometric optics has been used extensively to accurately model the RCS of targets at high carrier frequencies. However, the technique is computationally extensive and hence not suitable for modeling the time-varying RCS of dynamic human motions, at radar sampling frequencies, since humans are spatially large three-dimensional dielectric bodies with considerable variation in posture and pose. A computationally simpler alternative for modeling radar signatures of human motion is based on the scattering center model. However, the reflectivities of the scattering centers, in prior works, have been loosely approximated by RCS values of primitives resembling body shapes resulting in inaccurate estimates of RCS magnitudes. 

In our work, we hypothesize that the scattering coefficients fluctuate very slowly over multiple CPIs while the positions of the scatterers change rapidly across multiple PRIs. Therefore, we estimate the scattering center coefficients by combining the point scatterer model with the ray tracing RCS estimates in a linear regression framework. The positions of the scattering centers are obtained from an animation model of a pedestrian gathered from MoCap data. We use the reflectivity estimates to obtain realistic radar scattered signal that are processed to obtain commonly used radar signatures such as range-time, Doppler-time and range-Doppler ambiguity plots. Simultaneous to the MoCap data collection, we gathered measurement data using an automotive radar at 77 GHz from which the radar signatures were generated. The simulated signatures showed a low normalized mean square error (below 10\%) and high structural similarity (above 80\%) with respect to the measured signatures indicating the efficacy of the proposed method. We also demonstrated the versatility of our simulation method for modeling radar signatures at different polarizations, aspect angles and carrier frequencies.
\section{Appendix}
\subsection{Pedestrian RCS at alternate radar configurations}
\begin{figure*}[!ht]
    \centering
    \subfloat[\label{fig:Monostatic_front_24GHz}]{\includegraphics[width=0.3\textwidth]{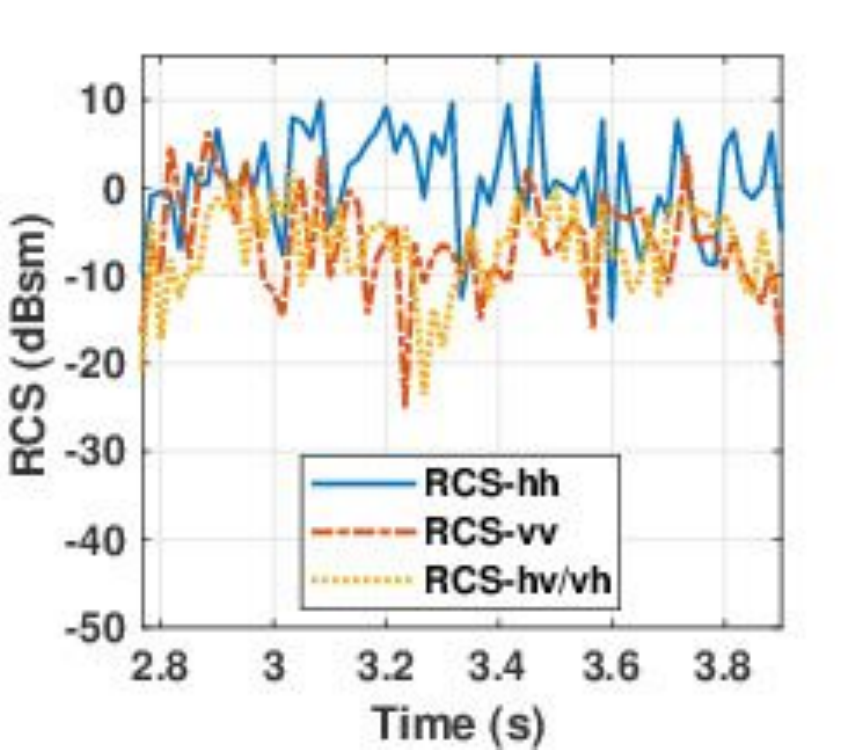}}\hfil
    \subfloat[\label{fig:Monostatic_45deg_24GHz}]{\includegraphics[width=0.3\textwidth]{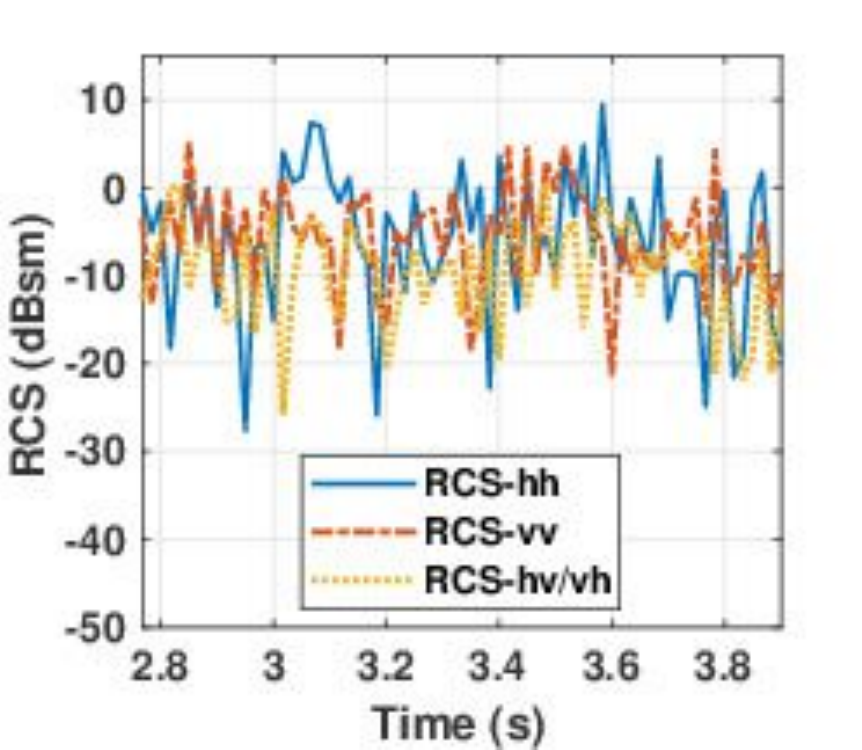}}\hfil
    \subfloat[\label{fig:Monostatic_90deg_24GHz}]{\includegraphics[width=0.3\textwidth]{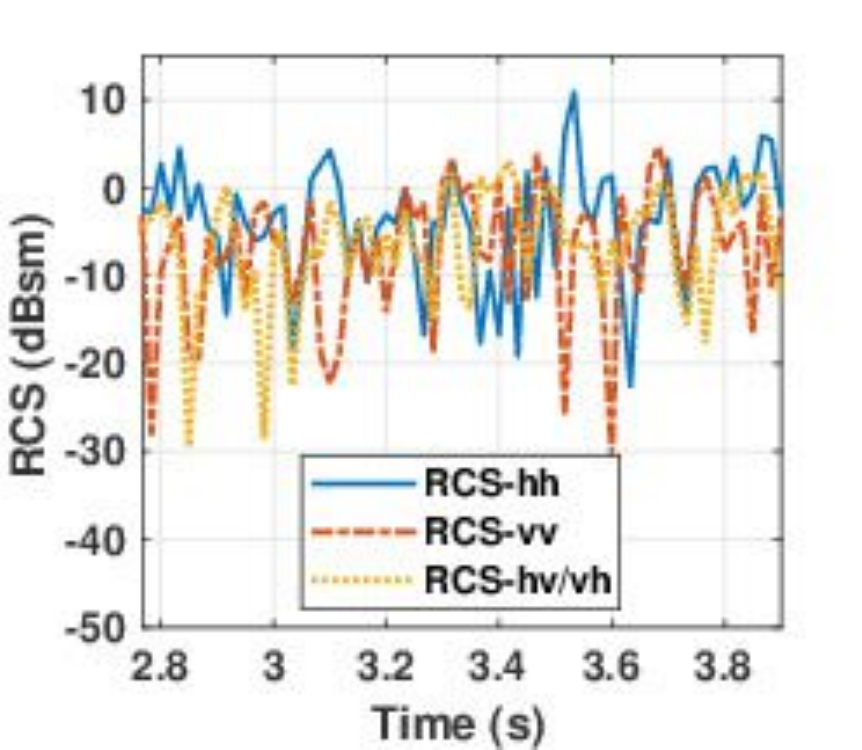}}
    \caption{Simulated monostatic ($\phi^{i} = \phi^s $) RCS across multiple frames corresponding to one walking stride obtained from ray tracing at 24 GHz for three aspect angles (a) front incidence ($\phi^{i} = 0^{\circ}$) (b) oblique incidence ($\phi^{i} = 45^{\circ}$) and (c) $90^{\circ}$ incidence ($\phi^{i} = 90^{\circ}$).}
    \label{fig:Monostatic_at_24}
\end{figure*}

\begin{figure*}[!ht]
    \centering
    \subfloat[\label{fig:Bistatic_front_77GHz_182}]{\includegraphics[width=0.3\textwidth]{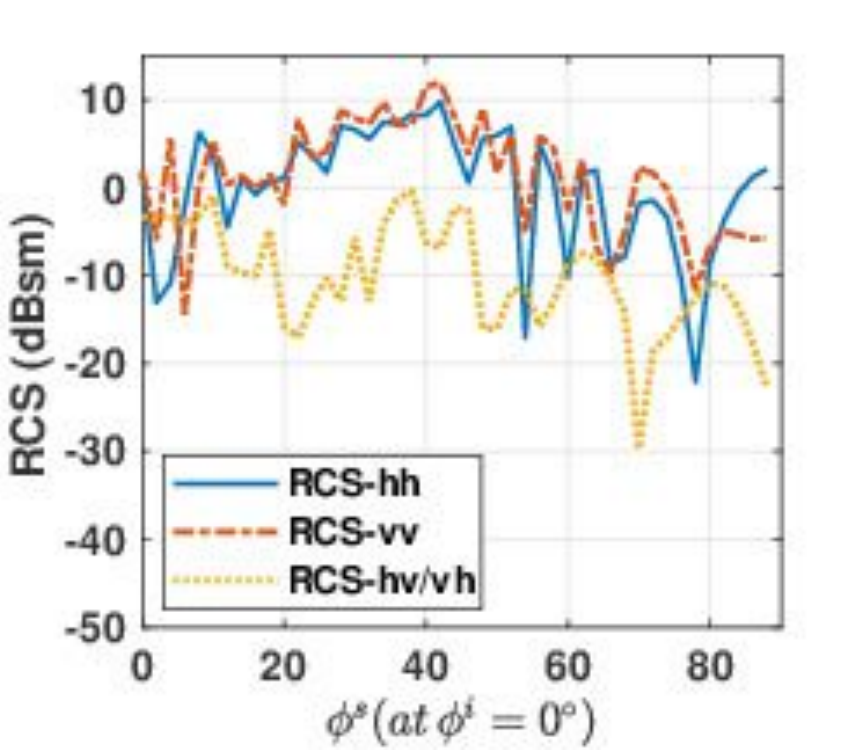}}\hfil
    \subfloat[\label{fig:Bistatic_45deg_77GHz_182}]{\includegraphics[width=0.3\textwidth]{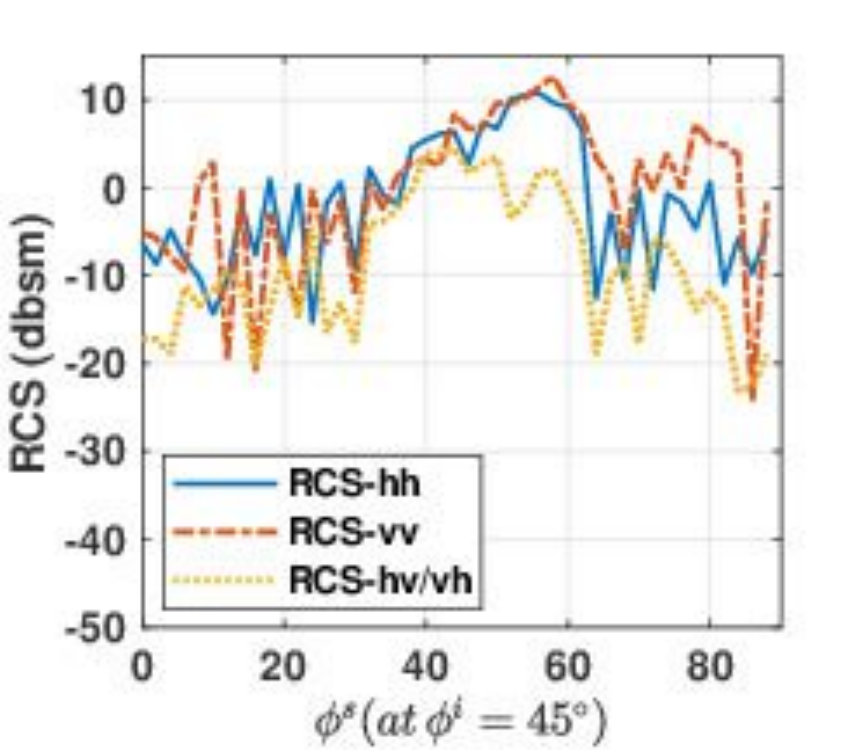}}\hfil
    \subfloat[\label{fig:Bistatic_90deg_77GHz_182}]{\includegraphics[width=0.3\textwidth]{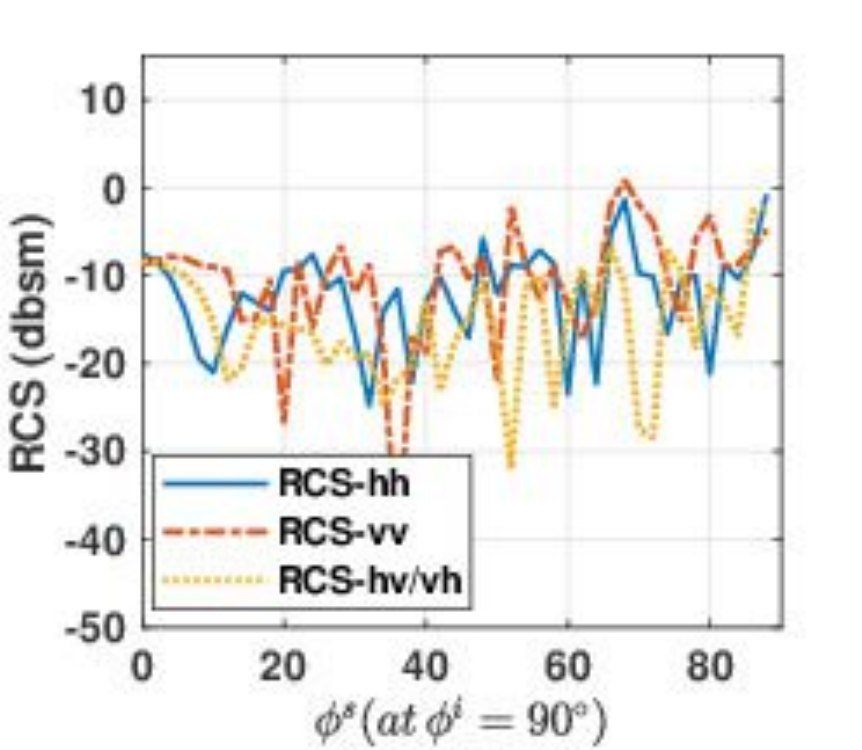}}\hfil
\caption{Simulated bistatic RCS versus $\phi^s$ such that bistatic aspect angle $= \phi^s - \phi^{i}$; for single frame/pose for three aspect angles (a) front incidence ($\phi^{i} = 0^{\circ}$) (b) oblique incidence ($\phi^{i} = 45^{\circ}$) and (c) $90^{\circ}$ incidence ($\phi^{i} = 90^{\circ}$) at 77 GHz.}
    \label{fig:Bistatic1}
\end{figure*}

\begin{figure*}[!ht]
    \centering
    \subfloat[\label{fig:Bistatic_front_24GHz_182}]{\includegraphics[width=0.3\textwidth]{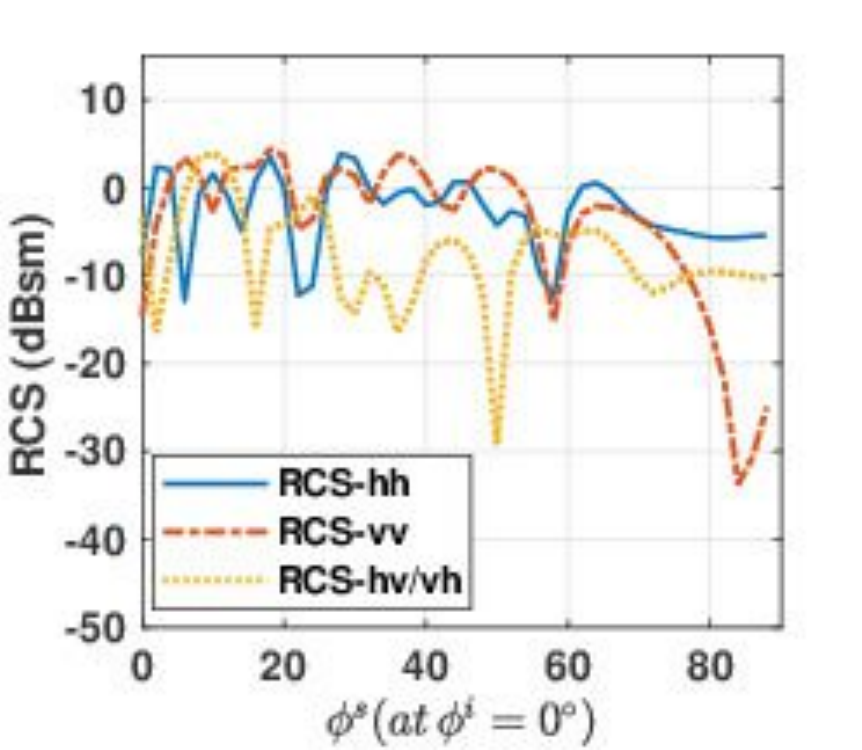}}\hfil
    \subfloat[\label{fig:Bistatic_45deg_24GHz_182}]{\includegraphics[width=0.3\textwidth]{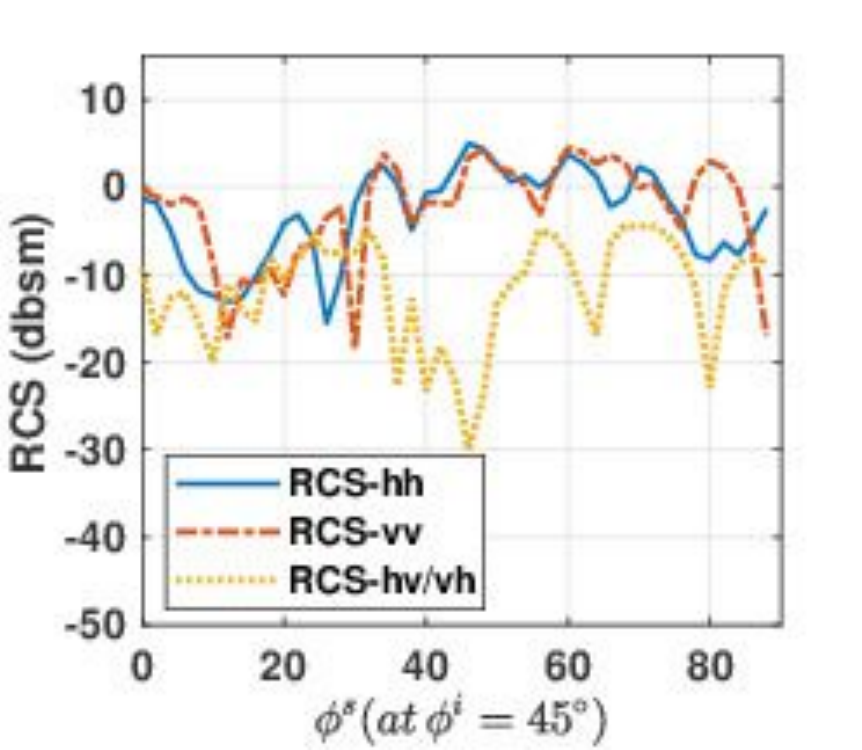}}\hfil
    \subfloat[\label{fig:Bistatic_90deg_24GHz_182}]{\includegraphics[width=0.3\textwidth]{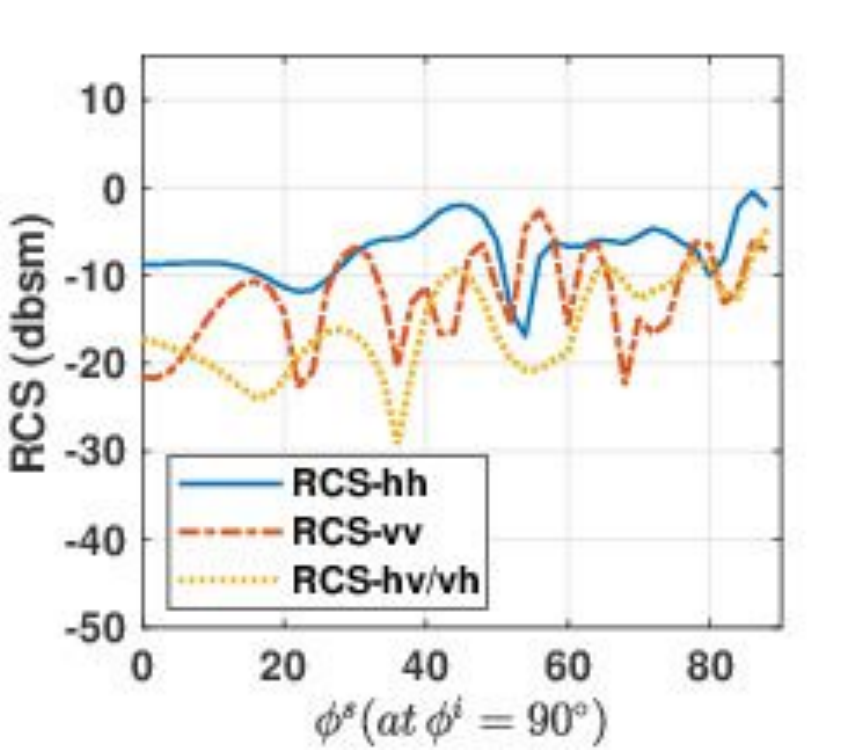}}
    \caption{Simulated bistatic RCS versus $\phi^s$ such that bistatic aspect angle $= \phi^s - \phi^{i}$; for single frame/pose for three aspect angles (a) front incidence ($\phi^{i} = 0^{\circ}$) (b) oblique incidence ($\phi^{i} = 45^{\circ}$) and (c) $90^{\circ}$ incidence ($\phi^{i} = 90^{\circ}$) at 24 GHz.}
    \label{fig:Bistatic2}
\end{figure*}

The second popular band of carrier frequencies for automotive radar is 24 GHz \cite{wenger2005automotive}. We present the monostatic RCS for different polarizations and incident angles in Fig.~\ref{fig:Monostatic_at_24}. The figure shows that the RCS values are slightly higher for the horizontal co-polarization scenario when compared to the vertical polarization especially for the case of frontal incidence ($0^{\circ}$). On average, the frontal incidence also gives rise to the highest RCS values for the monostatic configuration.
In some $V2X$ applications, it may be useful to have bistatic RCS of pedestrians. Fig.~\ref{fig:Bistatic1} and Fig~\ref{fig:Bistatic2}, present the variation of RCS with $\phi^s$ for bistatic angle $= \phi^s-\phi^{i}$; for a single frame/pose for different polarizations and for three different incident angles for two automotive frequencies 24 GHz and 77GHz. The bistatic RCS corresponds to the monostatic RCS when $\phi^s = \phi^{i}$. Interesting, the bistatic RCS is higher than the monostatic RCS at some aspect angles for some postures of the human. 
\subsection{Computational complexity of the proposed approach}
\begin{figure}[!ht]
 \centering
 \includegraphics[width = 0.27\textwidth]{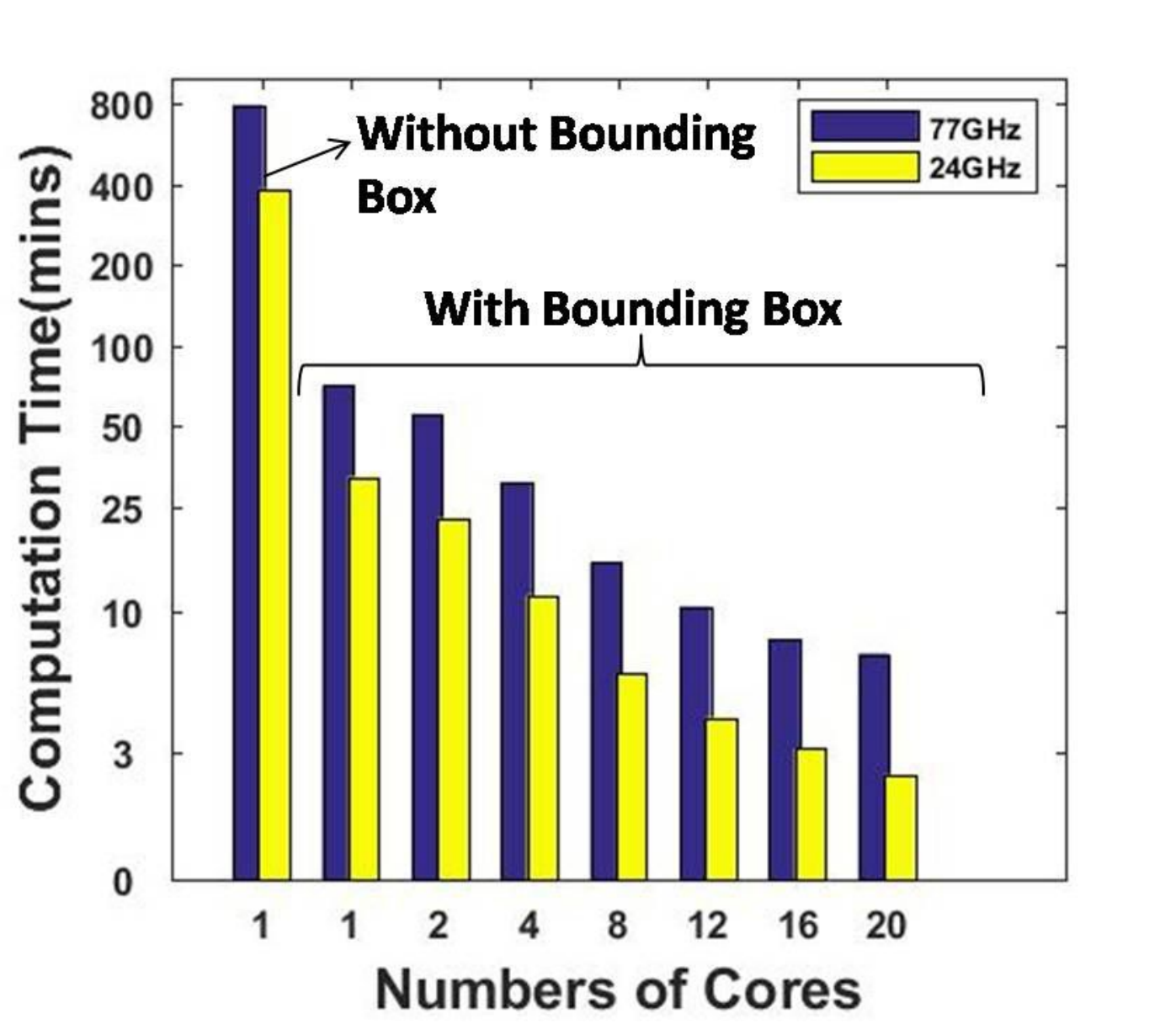}
     \caption{Reduction in computation time of RCS using electromagnetic ray tracing, with introduction of bounding box technique and increasing the number of cores for parallel processing at 77 GHz and 24 GHz}
     \label{fig: Bar_graph}
 \end{figure}

In Fig.~\ref{fig: Bar_graph}, we indicate the computational time for generating the results for different processing configurations. A realistic model of the human requires the body to be rendered by a large number of small sized triangular facets. For ray tracing, we consider a large volume of parallel illumination rays emanating from an illumination plane from grid points that must be densely placed at least $\frac{\lambda}{10}$ apart. The computational complexity is determined by the intersection tests between all the illumination rays and the facets on the body. This results in considerable complexity (800 minutes to compute RCS at 77 GHz in Fig.~\ref{fig: Bar_graph}). 
Several works in graphics have addressed the challenges of reducing the computational complexity associated with ray tracing \cite{rubin19803, fujimoto1985accelerated}. We have implemented the bounding box test in our work where the poly-mesh human is divided into several distinct parts each enclosed by a spatial bounding box. Instead of testing every ray with every triangle, we test every ray with every bounding box. Only if the ray intersects the bounding box, do we test the intersection of the ray with every facet within the bounding box. By using bounding box technique on a single core processor, we observed about 14 times reduction in computation time from 800 to 60 minutes in Fig.~\ref{fig: Bar_graph}.
Since the ray-triangle intersection tests can be carried out in parallel, the computation time can be further reduced by implementing the algorithm across multiple parallel processors. The algorithm was implemented using the parallel computing tool box of Matlab. 
By using the parallel processing with the bounding box technique, the computation time was further reduced to 8 minutes for a 20 core system.
\bibliographystyle{ieeetran}
\bibliography{references}

\end{document}